\documentclass[10pt,twocolumn,letterpaper]{article}

\usepackage[pagenumbers]{cvpr} 

\usepackage[dvipsnames]{xcolor}

\newcommand{\bI}{\mathbf{I}}

\newcommand{\bK}{\mathbf{K}}

\newcommand{\bM}{\mathbf{M}}

\newcommand{\bp}{\mathbf{p}}

\newcommand{\bR}{\mathbf{R}}

\newcommand{\bt}{\mathbf{t}}

\newcommand{\bepsilon}{\boldsymbol{\epsilon}}

\newcommand{\cF}{\mathcal{F}}

\newcommand{\cI}{\mathcal{I}}

\newcommand{\cL}{\mathcal{L}}

\newcommand{\cN}{\mathcal{N}}

\makeatletter
\DeclareRobustCommand\onedot{\futurelet\@let@token\@onedot}
\def\@onedot{\ifx\@let@token.\else.\null\fi\xspace}
\def\eg{e.g\onedot} 
\def\ie{i.e\onedot}

\makeatother

\definecolor{yellow}{rgb}{1, 1, 0.7}
\definecolor{orange}{rgb}{1, 0.85, 0.7}
\definecolor{tablered}{rgb}{1, 0.7, 0.7}
\definecolor{red}{rgb}{1, 0, 0}

\definecolor{wincolor}{rgb}{0.85, 0.0, 0.0}

\definecolor{darkyellow}{rgb}{0.8, 0.8, 0.5}
\definecolor{darkred}{rgb}{0.7, 0.3, 0.3}
\definecolor{darkgreen}{rgb}{0.3, 0.7, 0.3}
\definecolor{blue}{rgb}{0.251, 0.498, 0.824}
\definecolor{green}{rgb}{0, 1.0, 0}
\definecolor{pink}{rgb}{1, 0.4, 0.7}

\newcommand{\boldparagraph}[1]{\vspace{0.1cm}\noindent{\bf #1:}}

\definecolor{cvprblue}{rgb}{0.21,0.49,0.74}
\usepackage[pagebackref,breaklinks,colorlinks,allcolors=cvprblue]{hyperref}

\usepackage{pifont}
\usepackage{color}
\usepackage{xcolor}
\usepackage{nicefrac}       %
\usepackage{multirow}
\usepackage{multicol}
\usepackage{dsfont}

\usepackage{colortbl}
\usepackage[percent]{overpic}
\usepackage[accsupp]{axessibility}

\newcommand{\co}[1]{\textcolor{red}{#1}}

\newcommand{\zerodisplayskips[1]}{
    \setlength{\abovedisplayskip}{#1pt}
    \setlength{\belowdisplayskip}{#1pt}
    \setlength{\abovedisplayshortskip}{#1pt}
    \setlength{\belowdisplayshortskip}{#1pt}
}

\def\paperID{6250} 
\def\confName{CVPR}
\def\confYear{2025}

\title{EVPGS: Enhanced View Prior Guidance for Splatting-based \\Extrapolated View Synthesis}

\author{Jiahe Li
\and
Feiyu Wang$^*$
\and
Xiaochao Qu
\and
Chengjing Wu
\and
Luoqi Liu$^*$
\and
Ting Liu$^*$
\and
MT Lab, Meitu Inc., Beijing 100083, China
\\
\small \texttt{\{ljh15, wfy1, qxc, ethan, llq5, lt\}@meitu.com}\\
\small $^*$Corresponding Author
}

\begin{document}
\maketitle
\begin{abstract}
Gaussian Splatting (GS)-based methods rely on sufficient training view coverage and perform synthesis on interpolated views. In this work, we tackle the more challenging and underexplored Extrapolated View Synthesis (EVS) task. Here we enable GS-based models trained with limited view coverage to generalize well to extrapolated views. To achieve our goal, we propose a view augmentation framework to guide training through a coarse-to-fine process. At the coarse stage, we reduce rendering artifacts due to insufficient view coverage by introducing a regularization strategy at both appearance and geometry levels. At the fine stage, we generate reliable view priors to provide further training guidance. To this end, we incorporate an occlusion awareness into the view prior generation process, and refine the view priors with the aid of coarse stage output. We call our framework Enhanced View Prior Guidance for Splatting (EVPGS). To comprehensively evaluate EVPGS on the EVS task, we collect a real-world dataset called Merchandise3D dedicated to the EVS scenario. Experiments on three datasets including both real and synthetic demonstrate EVPGS achieves state-of-the-art performance, while improving synthesis quality at extrapolated views for GS-based methods both qualitatively and quantitatively. 
Our code and dataset are available on the  \href{https://charley077.github.io/EVPGS_Homepage/}{EVPGS Homepage}.
\end{abstract}  
\section{Introduction}
\label{sec:intro}

\begin{figure}
\centering
\includegraphics[width=1.0\linewidth]{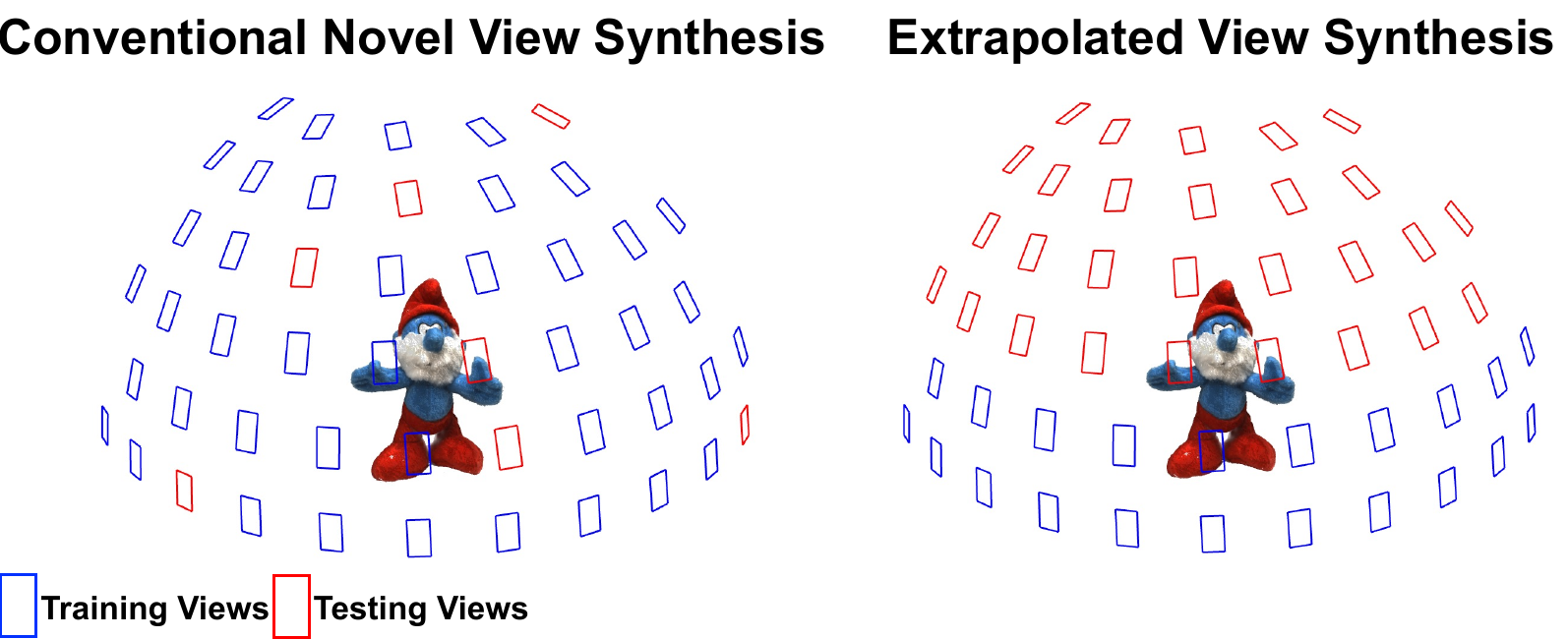}
\caption{\textbf{Illustration of the EVS problem.} We visualize the camera pose at each view as a rectangle warped by the pose. \textbf{Left:} In conventional novel view synthesis, the training views sufficiently cover the scene. \textbf{Right:} In EVS, the training views have limited scene coverage (only horizontal and near-horizontal views as per the example) and overlap poorly with the testing views. Example scene from DTU~\cite{jensen2014large}.}
\vspace{-15pt}
\label{fig:evs}
\end{figure}

Gaussian Splatting (GS)~\cite{kerbl20233dgaussiansplattingrealtime} and its variants~\cite{yu2024mip-splatting,mallick2024taming, kerbl2024hierarchical, radl2024stopthepop} have enabled high-fidelity and real-time novel view synthesis. Despite the achievements, a main drawback of the existing methods is that they rely on training images with sufficient view coverage, which is an interpolated view synthesis setting.
In real-world scenarios, users often capture images by moving the camera horizontally around an object, yet may desire novel views from elevated pitch angles. Therefore, existing methods suffer from deteriorated synthesis quality in real-world applications due to both limited training view coverage and the resulting large disparity between the training and testing views. This gives rise to the extrapolated view synthesis (EVS) problem, which we tackle in this work. We illustrate EVS in Figure~\ref{fig:evs}.

\begin{figure}
\centering
\includegraphics[width=1.0\linewidth]{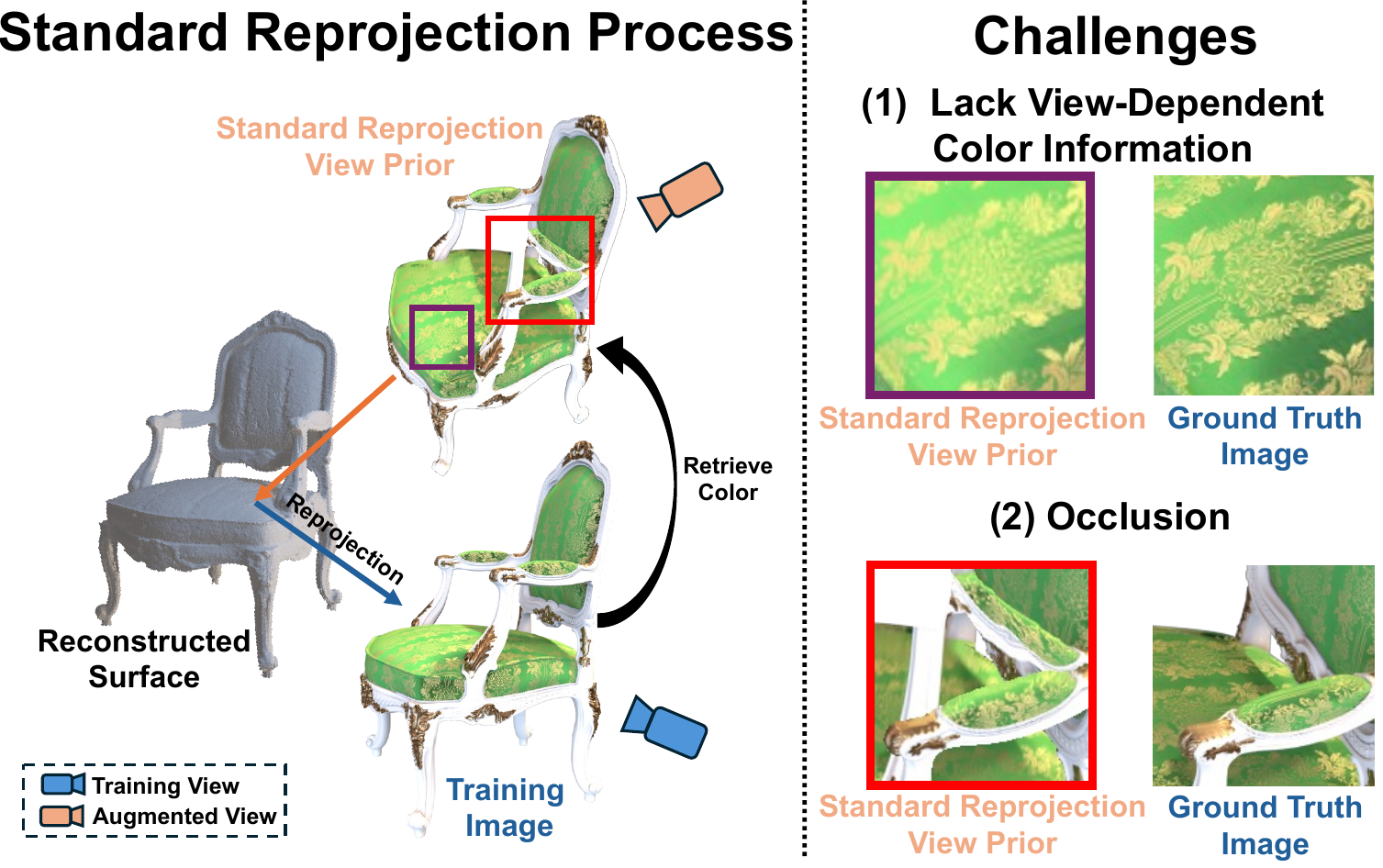}
\vspace{-10pt}
\caption{
\textbf{Illustration of the standard reprojection process and its challenges.} 
\textbf{Left:} In the standard reprojection process, each pixel in the view prior of an augmented view retrieves color from the corresponding pixel in the nearest training view, using reprojection via the reconstructed mesh. 
\textbf{Right:} This technique risks losing view-dependent color information for the augmented views and may introduce occlusion artifacts, leading to corrupted view priors.
}
\vspace{-15pt}
\label{fig:challenge}
\end{figure}

To remedy for the gap between the training and testing views, concurrent works
in sparse view reconstruction~\cite{xu2024mvpgsexcavatingmultiviewpriors, liu2024deceptive,yu2024lmgaussianboostsparseview3d} perform view augmentation by generating view priors at augmented views, which act as pseudo-labels during training.
The view prior is typically generated by reprojecting a training-view image to an unseen augmented view 
via a reconstructed explicit 3D representation, \eg mesh or depth map. While these sparse view reconstruction methods can produce novel-view renderings 
with reasonable realism, they essentially deal with interpolated view synthesis.
We scrutinise the EVS problem and identify two challenges which have not been addressed by existing strategies for generating view priors, \ie lack of view-dependent color information, and occlusions in complex scenes. We illustrate these challenges in Figure~\ref{fig:challenge}. These issues can cause rendering artifacts such as unrealistic appearance and corrupted scene parts, which call for more dedicated processing of the view priors.

To address both challenges above, we present a comprehensive view augmentation framework for EVS called \textbf{E}nhanced \textbf{V}iew \textbf{P}rior \textbf{G}uidance for \textbf{S}platting (EVPGS).
Our framework generates reliable view priors through a two-stage coarse-to-fine process. At the coarse stage, we leverage both the prior knowledge from the Denoising Diffusion Model (DDM)~\cite{ho2020denoisingdiffusionprobabilisticmodels, rombach2022highresolutionimagesynthesislatent} and the reconstructed mesh from pre-trained GS Model to  
reduce rendering artifacts at extrapolated views caused by insufficient view coverage.
At the fine stage, we incorporate the occlusion-aware strategy into the view prior generation to eliminate the occlusion issue encountered by the standard reprojection technique (illustrated in Figure~\ref{fig:challenge}). We also refine the view priors
by integrating the view-dependent color information from the regularized renderings at the coarse stage. Our fine stage
can produce high-quality view priors as strong appearance guidance towards synthesizing realistic extrapolated views.
We term the view priors generated by the EVPGS framework \textit{Enhanced View Priors}.

To comprehensively evaluate our EVPGS framework, we collect \textit{Merchandise3D}, a real-world dataset specifically tailored for the EVS task. 
This dataset includes seven object instances of several categories with intricate structural and textural details, representing challenging real-world scenarios.
We evaluate EVPGS on three different benchmark datasets, both real and synthetic, including DTU~\cite{jensen2014large}, Synthetic-NeRF~\cite{mildenhall2020nerfrepresentingscenesneural}, and our Merchandise3D. Experimental results demonstrate the superior performance of EVPGS over the different datasets.

In summary, our main contributions are as follows,
\begin{enumerate}
    \item  
    To address the challenges posed by EVS, we propose the view augmentation framework EVPGS based on GS models to generate reliable view priors as high-quality augmented views.
    \item 
    We propose several complementary strategies to improve view prior quality through a coarse-to-fine process. At the coarse stage, we introduce a regularization strategy at both appearance and geometry levels to reduce rendering artifacts due to insufficient training view coverage. Then at the fine stage, we generate reliable view priors by both eliminating the occlusion issue in the standard reprojection technique, and refining the view prior with view specific visual information.
    \item EVPGS can synthesize extrapolated views with realistic appearance and fine details. Experiments demonstrate EVPGS achieves state-of-the-art performance on three EVS benchmarks, 
    including both public and our self-collected datasets. We release our Merchandise3D dataset to facilitate future research.
\end{enumerate}
\section{Related Work}
\label{sec:related_work}

\subsection{Neural Scene Representations}
\label{sec:neural_implicit_representation}
Neural scene representations have been extensively studied for efficient and high-quality 3D reconstruction. The pioneering work Neural Radiance Fields (NeRF) \cite{mildenhall2020nerfrepresentingscenesneural,barron2021mip} utilizes a deep Multi-layer Perceptron to encode volume density and color across the continuous 3D space and employs the alpha-composition technique for rendering into 2D images. \cite{wang2021neus,wang2023neus2,yariv2021volume} introduces explicit surface representation for better reconstruction. 
Several other works~\cite{hedman2021snerg,yu2021plenoctrees,yu2022plenoxels,mueller2022instant-ngp,wu2023voxurf,fridovichkeil2023kplanesexplicitradiancefields} adopt the voxel-grid representation to expedite training and rendering.
To reduce memory cost, \cite{chen2022tensorftensorialradiancefields, fridovichkeil2023kplanesexplicitradiancefields} decompose the 3D volume into a set of 2D/1D planes/vectors, representing the color and opacity at each 3D point with features stored in the planes/vectors.

A more recent line of works~\cite{huang20242d, kerbl20233dgaussiansplattingrealtime, Yu2024GOF, yu2024mip-splatting, zhang2024radegsrasterizingdepthgaussian, guedon2024sugar, mallick2024taming, kerbl2024hierarchical, radl2024stopthepop} adopt the splatting technique~\cite{zwicker2001ewa} for rendering, achieving significant improvement in both efficiency and quality. The seminal work 3D Gaussian Splatting (3DGS) \cite{kerbl20233dgaussiansplattingrealtime} proposes the Gaussian primitive representation which is flexible and parameter-efficient while being conducive to parallelized implementation. Mip-Splatting \cite{yu2024mip-splatting} introduces kernel filtering techniques to alleviate aliasing artifacts caused by resolution changes. By exploiting the characteristics of Gaussian primitives, \cite{zhang2024radegsrasterizingdepthgaussian, huang20242d, Yu2024GOF, guedon2024sugar} design dedicated approaches to improve surface reconstruction. 
Our framework is generic and can be implemented with different GS-based methods as backbones.

\subsection{View Augmentation}\label{Sec:EVS}
\boldparagraph{Sparse View Reconstruction} While sparse view reconstruction~\cite{chen2024mvsplat,charatan2024pixelsplat3dgaussiansplats,niemeyer2021regnerfregularizingneuralradiance,wang2024freesplat,yang2023freenerfimprovingfewshotneural, wang2023sparsenerfdistillingdepthranking, li2024dngaussian, paliwal2025coherentgs, zhu2025fsgs,liu2024deceptive, xu2024mvpgsexcavatingmultiviewpriors, yu2024lmgaussianboostsparseview3d} aims to improve view interpolation between a small set of training views, several works~\cite{liu2024deceptive, xu2024mvpgsexcavatingmultiviewpriors, yu2024lmgaussianboostsparseview3d, zhu2025fsgs} incorporate view augmentation strategies. \cite{liu2024deceptive, xu2024mvpgsexcavatingmultiviewpriors} leverage the reprojection technique to generate view priors, which provide training supervision at additional views. \cite{yu2024lmgaussianboostsparseview3d} improves the geometric consistency across different augmented views by exploiting the point cloud generated with 
Structure-from-Motion (SfM)~\cite{schonberger2016structure,sarlin2019coarse}. \cite{zhu2025fsgs} leverages augmented views 
to encourage the newly cloned Gaussians to be distributed around the more representative locations.

\boldparagraph{Extrapolated View Synthesis} Several other works 
utilize view augmentation strategies to better synthesize extrapolated views~\cite{yang2023nerfvs, zhang2022raypriors, hwang2024vegsviewextrapolationurban}. 
Inspired by the early method based on depth probability volume~\cite{choi2019extreme}, \cite{yang2023nerfvs} utilizes the reconstructed pseudo-depth map with additional view coverage information to produce priors at extrapolated views with reasonable fidelity. \cite{zhang2022raypriors} leverages the reprojection technique to apply supervision on a set of augmented pixels based on reconstructed mesh vertices. Different from the methods using augmented views for pixel-wise supervision, \cite{hwang2024vegsviewextrapolationurban} proposes an DDM-based~\cite{rombach2022highresolutionimagesynthesislatent} 
regularization strategy to remove artifacts 
caused by insufficient training view coverage at the image semantics level. 
In our framework, we encourage the backbone to extrapolate to unseen views by adopting a similar strategy to \cite{hwang2024vegsviewextrapolationurban}, while we additionally consider both problems of lacking view-dependent color information and
occlusion issue during view prior generation for further improving fidelity of the view prior.
\section{Method}

\begin{figure*}
    \centering
    \includegraphics[width=1.0\linewidth]{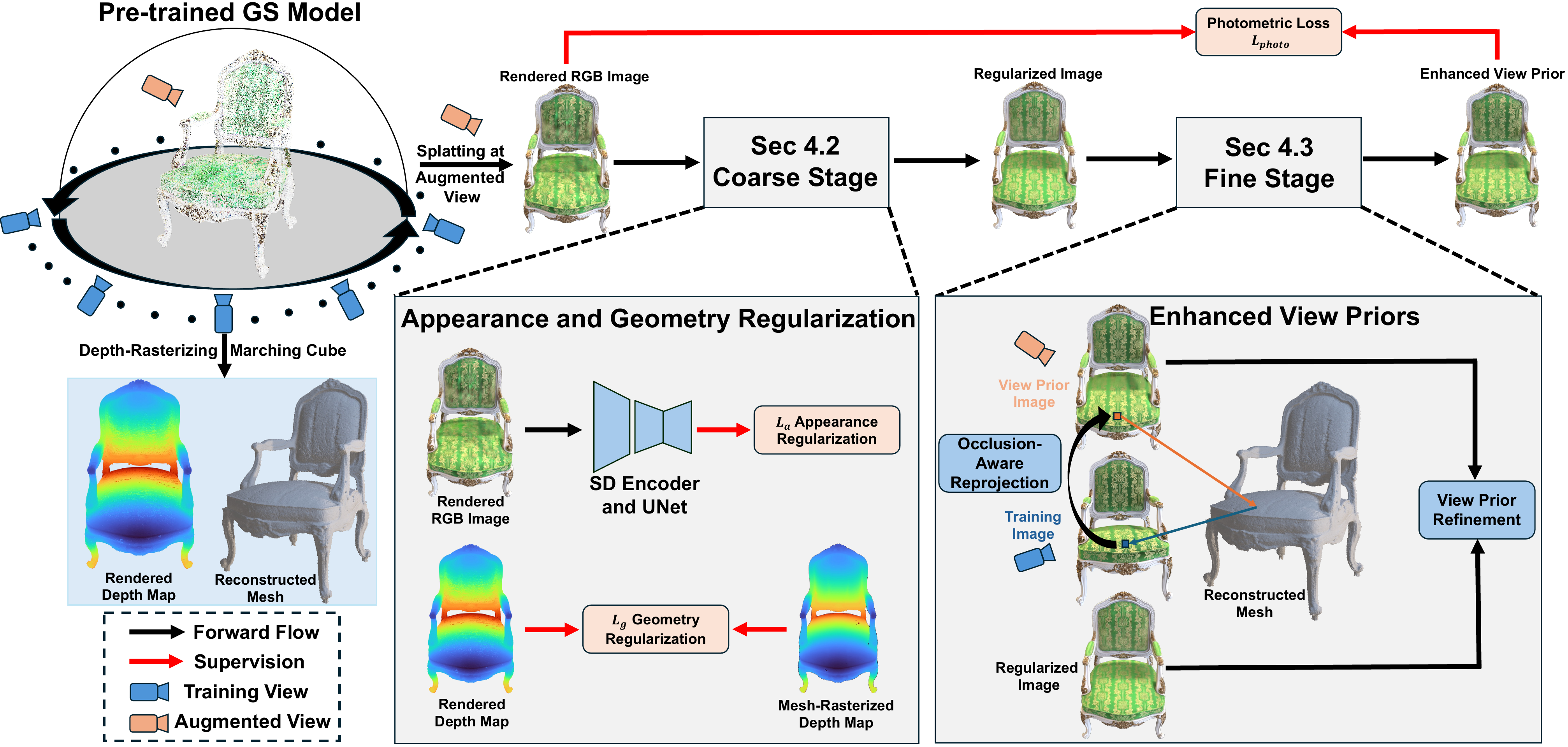}
    \caption{
    \textbf{Framework Overview.} In EVPGS, we first pre-train a GS model using the training set with limited view coverage (\eg only horizontal views), then fine-tune the GS model at augmented views (\eg obtained by elevating the training views) through a coarse-to-fine process. At the coarse stage, we propose the Appearance and Geometry Regularization (AGR) strategy, where we use the pre-trained GS model to produce synthesis at augmented views and reduce the artifacts at these views by leveraging the Denoising Diffusion Model ~\cite{rombach2022highresolutionimagesynthesislatent}. We additionally generate depth maps rasterized from reconstructed mesh to supervise the depth maps directly rendered from the GS model. At the fine stage, we produce \textit{Enhanced View Priors} at these augmented views as pseudo-labels, via our Occlusion-Aware Reprojection and Refinement (OARR) strategy. The OARR strategy comprises both the occlusion-aware reprojection technique to eliminate occlusion corruptions, and the view prior refinement strategy to incorporate the view-dependent colors obtained from the coarse stage.
    }
    \label{fig:pipeline}
    \vspace{-15pt}
\end{figure*}
\subsection{The EVPGS Framework}\label{Sec:pipeline}
In the EVS task, the goal is to synthesize extrapolated views by leveraging a training dataset with only limited view coverage. As shown in Figure \ref{fig:evs}, there can be a drastic discrepancy between the training and testing view coverages, which hampers the learned GS model from synthesizing high-quality images at testing views.

Our key idea is based on exploiting view priors as augmented views during training. We propose the EVPGS framework for generating reliable view priors which we term \textit{Enhanced View Priors}. Our EVPGS framework conducts artifact removal and appearance refinement for GS models in a coarse-to-fine manner. We illustrate our overall framework in Figure~\ref{fig:pipeline}.

We first pre-train a GS model using the training views with limited coverage. Then at the coarse stage, we propose the Appearance and Geometry Regularization (AGR) strategy to fine-tune the pre-trained model across the augmented views. The regularization objective is designed by leveraging both a pre-trained DDM~\cite{rombach2022highresolutionimagesynthesislatent} and the reconstructed mesh from the pre-trained GS model. Benefiting from the strong appearance priors in DDM, our appearance regularization can produce visually sanitized images with significantly reduced artifacts. To further reduce artifacts, we additionally propose the geometry regularization strategy which indirectly improves appearance by smoothening the reconstructed surface at the augmented views. This is implemented by supervising the directly rendered depth map from pre-trained GS model through mesh-rasterized depth map from the reconstructed mesh.

At the fine stage, we guide EVS towards both better realism and finer details by introducing the Occlusion-Aware Reprojection and Refinement (OARR) strategy, which generates \textit{Enhanced View Priors} through both occlusion removal and view prior refinement. To eliminate occlusion during view prior generation, we extend the standard reprojection technique~\cite{xu2024mvpgsexcavatingmultiviewpriors, liu2024deceptive, zhang2022raypriors} by introducing the reverse-reprojection process, and 
replace the occlusion-corrupted pixels with the coarse stage results
by designing a selection criterion based on reverse-reprojection. As a result of reprojection, the view prior only contains visual information from its corresponding training image and lacks the color information specific to its view. To compensate for the view-specific visual information, we introduce a simple but effective refinement strategy to incorporate the complementary visual information from coarse stage renderings into the view priors.

While each stage of EVPGS can achieve significant improvement individually, we combine both stages into one comprehensive framework for the best performance. We present our proposed components in detail below.

\subsection{Appearance and Geometry Regularization}\label{Sec:AGR}
\boldparagraph{Appearance Regularization} 
At the coarse stage, we fine-tune the pre-trained GS model augmented views to reduce artifacts due to insufficient training view coverage. To achieve this goal, we employ the estimated noise of Denoising Diffusion Model (DDM)~\cite{rombach2022highresolutionimagesynthesislatent} as the evaluation criterion for the augmented views.

For each rendered image $\hat{\cI}$ at an augmented view, the DDM produces the noise estimate through a diffusion and denoising process. In the diffusion process, DDM gradually adds random Gaussian noise $\bepsilon \sim \cN(\mathbf{0},\bI)$ to $\hat{\cI}$ following the process $\hat{\cI}_{\tau}=\sqrt{\bar{\alpha}_{\tau}}\hat{\cI}+(1-\bar{\alpha}_{\tau})\bepsilon$, where $\hat{\cI}_{\tau}$ is the noisy version of $\hat{\cI}$, $\tau$ is the time step~\cite{ho2020denoisingdiffusionprobabilisticmodels} and $\bar{\alpha}_{\tau}$ is the pre-defined noise variance schedule. Then in the denoising process, the noise estimatation is output by the denoising model $\epsilon_{\mathbb\theta}(\cdot, \tau)$, where $\theta$ is the model parameters. Previous works~\cite{ho2020denoisingdiffusionprobabilisticmodels, vincent2011connection} have shown the estimated noise of DDM is proportional to the score function, which is the negative gradient of the log-likelihood of $\hat{\cI}$ in our scenario, \ie,
\begin{equation}
\zerodisplayskips[1]
    \epsilon_{\theta}(\hat{\cI}_{\tau},\tau) \propto -\nabla_{\hat{\cI}}\log{p(\hat{\cI})}
\end{equation}
This equation suggests that by computing the estimated noise, we can obtain the direction of the image gradient $\nabla_{\hat{\cI}}\log{p(\hat{\cI})}$ which drives $\hat{\cI}$ towards the data distribution modes captured by DDM. Therefore, as suggested in \cite{wynn2023diffusionerfregularizingneuralradiance, hwang2024vegsviewextrapolationurban}, we formulate our appearance regularization term $\cL_{a}$ as follows\footnote{Following the derivation of~\cite{wynn2023diffusionerfregularizingneuralradiance}, the term $\epsilon_{\theta}(\hat{\cI}_{\tau},\tau)$ in Eq.~\ref{eq:loss_reg_geometry} does not receive gradient when optimizing the GS model.}, 
\begin{equation}\label{eq:loss_reg_geometry}
\zerodisplayskips[1]
 \cL_{a}=\sum(\epsilon_{\theta}(\hat{\cI}_{\tau},\tau) \odot \hat{\cI}_{\tau})
\end{equation}
where $\odot$ is the Hadamard product. 
Unlike previous works~\cite{wynn2023diffusionerfregularizingneuralradiance, hwang2024vegsviewextrapolationurban} which fine-tune the DDM model to adapt to their scenarios, we avoid fine-tuning the DDM model, thereby reducing computational costs and mitigating the risk of catastrophic forgetting.

\boldparagraph{Geometry Regularization} 
In the coarse stage of our framework, we observe the depth maps rendered from the pre-trained GS model (\ie denoted as $\hat{D}$) at augmented views are usually inaccurate,
which deteriorates the rendering results at the extrapolated views.
As a result, we use the reconstructed mesh from the pre-trained model to smooth the rendered depth map, further reducing severe artifacts.
To smooth the rendered depth map, for each augmented view, we use the pseudo-depth map (\ie denoted as $\bar{D}$) rasterized from the reconstructed mesh to enforce geometry regularization, which we formulate as supervision on the rendered depth map $\hat{D}$, \ie,
\begin{equation}\label{eq:loss_reg_appearance}
\zerodisplayskips[1]
    \cL_{g} = \frac{1}{H}\frac{1}{W}\sum_{i=1}^{H}\sum_{j=1}^{W}|\hat{D}_{ij}-\bar{D}_{ij}|
\end{equation}
where $H$ and $W$ are the image height and width.

\subsection{Enhanced View Priors}
\label{Sec:RRP}
At the fine stage, we generate reliable view priors to further improve realism at augmented views while retaining both the view-dependent effects inherit from the coarse stage and the high-frequency details which can be lost after the coarse stage. This is achieved by our joint strategy to both remove occlusion corruption and refine the view priors. We term this strategy 
Occlusion-Aware Reprojection and Refinement (OARR) strategy.

\boldparagraph{Occlusion-Aware Reprojection} We first generate view priors based on our improved reprojection technique. 
For each augmented view $v$, we rasterize the
depth map $\bar{D}_v$ at this augmented view and the depth map $\bar{D}_t$ at the nearest training view $t$ using reconstructed mesh from the pre-trained GS model.
For each pixel coordinate $\bp_{v}\in\mathbb{R}^{2}$ in the augmented view, the reprojection technique calculates its correspondence pixel $\bp_{v\rightarrow t}\in\mathbb{R}^{2}$ in the nearest training view, \ie,
\begin{equation}\label{Eq:reprojecion}
\zerodisplayskips[1]
    d_{\bp_{v\rightarrow t}}\left ( \begin{matrix}
\bp_{v\rightarrow t} \\
1 \\
\end{matrix} \right )
    =\bK_{t}(d_{\bp_{v}}\bR \bK_{v}^{-1} \left ( \begin{matrix}
\bp_{v} \\
1 \\
\end{matrix} \right )+\bt)
\end{equation}
where $\bK_{v}$ and $\bK_{t}$ are the camera intrinsic matrices from the views $v$ and $t$, 
$\bR$ and $\bt$ are the rotation matrix and translation vector both mapping from the camera space of the augmented view to the camera space of the training view, 
$d_{\bp_{v}}$ and $d_{\bp_{v\rightarrow t}}$ are the depths of $\bp_{v}$ and $\bp_{v\rightarrow t}$ from $\bar{D}_{v}$ and $\bar{D}_{t}$, respectively. 
Once the correspondence $\bp_{v\rightarrow t}$ is determined, the color at $\bp_{v}$ in the augmented view can be obtained by passing the color at $\bp_{v\rightarrow t}$ in the training view to $\bp_{v}$. Since $\bp_{v\rightarrow t}$ is usually a sub-pixel coordinate, we calculate its color using bilinear interpolation based on its four nearest pixels.

To avoid potential occlusion when generating view priors, we extend the standard reprojection technique by introducing the reverse-reprojection technique which eliminates occlusion corruptions. Our technique finds the reverse-correspondence in the augmented view for each $\bp_{v\rightarrow t}$ following a reverse manner to Eq.~\ref{Eq:reprojecion}, \ie,
\begin{equation}
\zerodisplayskips[1]
    d_{\bp_{t\rightarrow v}}\left ( \begin{matrix}
\bp_{t\rightarrow v} \\
1 \\
\end{matrix} \right )
    =\bK_{v}(d_{\bp_{v\rightarrow t}}\bR^{T} \bK_{t}^{-1} \left ( \begin{matrix}
\bp_{v\rightarrow t} \\
1 \\
\end{matrix} \right )-\bR^{T}\bt)
\label{Eq:reverse_reprojecion}
\end{equation}
where $\bp_{t\rightarrow v}$ is the reverse-correspondence of $\bp_{v\rightarrow t}$ and $d_{\bp_{t\rightarrow v}}$ is the depth of $\bp_{t\rightarrow v}$. We observe the pixel distance $e=\left\|\bp_{v}-\bp_{t\rightarrow v}\right\|_{2}$, which we call reverse-reprojection error, can be used as a criterion for detecting occlusion, \ie the larger the distance, the more likely $\bp_{v}$ is corrupted by occlusion. Therefore, we set an error threshold $\xi$ on $e$ to filter out the pixels that are likely corrupted. When $e<\xi$, we pass the color of $\bp_{v\rightarrow t}$ to $\bp_{v}$; otherwise $\bp_{v}$ adopts the corresponding pixel color of the regularized renderings from the coarse stage.

\boldparagraph{View Prior Refinement}\label{sec:fusion}
As a result of reprojection, each view prior is only a warped version of its corresponding training image and lacks visual information specific to its view.
This may lead to unrealistic appearance at the extrapolated views. To incorporate view-dependent information into the view priors, we refine the view priors by blending them with regularized renderings from the coarse stage in a complementary way. We observe that the view priors can better retain high-frequency details, while the regularized renderings from the coarse stage exhibit more view-dependent coloring. Therefore, for each view prior $\bar\cI$, we obtain the regularized rendering $\Tilde\cI$ from the coarse stage at the same augmented view, then blend the high-frequency component of $\bar\cI$ with the low-frequency component of $\Tilde\cI$ through weighted summation in the frequency domain. We update $\bar\cI$ to \textit{Enhanced View Prior} as follows,
\begin{equation}
\zerodisplayskips[1]
    \bar\cI \leftarrow \cF^{-1}\big(
        \bM \odot \cF(\bar\cI) + (1-\bM) \odot \cF(\Tilde\cI) 
    \big)
\label{Eq:refinement}
\end{equation}
where $\bM$ is the weighting map in the frequency domain,
$\cF$ and $\cF^{-1}$ are the Fast Fourier Transform~\cite{gonzalez2008} and its inverse transform, respectively. 
The weight map $\bM$ assigns a large weight $w_{h}$ to the center pixel, which corresponds to the high-frequency area, and a small weight $w_{l}$ to the outer edges, which correspond to the low-frequency area. 
The weight at each coordinate on $\bM$ decreases linearly from $w_h$ to $w_l$ as the distance between the coordinate and the center of $\bM$ (\ie the highest frequency) increases.

\begin{figure*}
    \centering
    \includegraphics[width=1.0\linewidth]{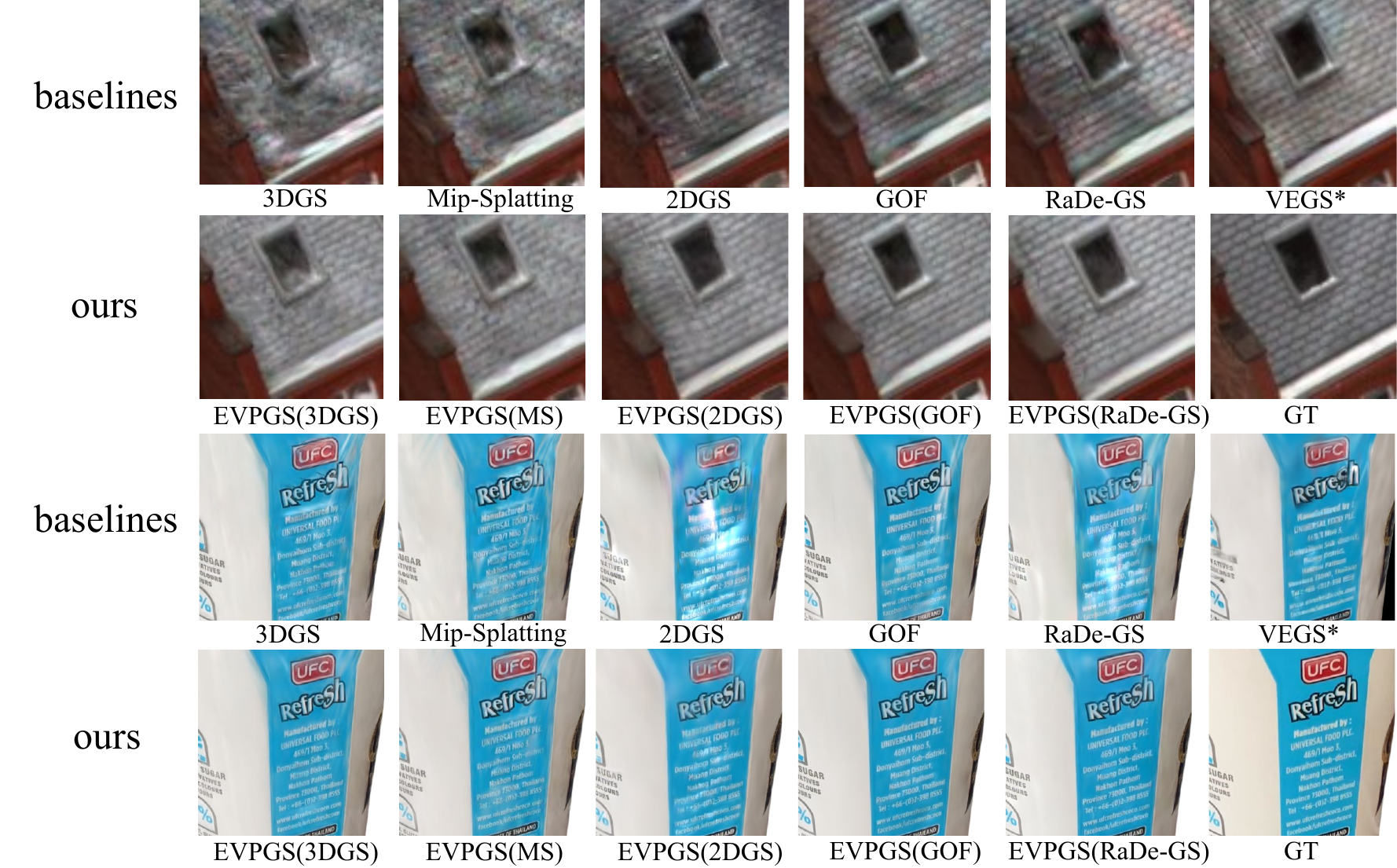}
    \caption{\textbf{Qualitative results on the real-world datasets DTU~\cite{jensen2014large} (top two rows) and our Merchandise3D (bottom two rows).} On both datasets, the baseline methods (the first and the third rows) exhibit rendering artifacts and lack of details due to limited training view coverage, while all five variants of EVPGS can effectively deal with these issues. 
    }
    \vspace{-10pt}
    \label{fig:real_compasison}
\end{figure*}

\subsection{Training Strategies}\label{Sec:train}

\boldparagraph{Pre-train Stage} During the pre-train stage, we use the photometric loss employed in GS-based methods, 
which is a combination of $L_{1}$ and SSIM\cite{wang2004image} loss, 
\ie,
\begin{equation}\label{eq:loss_photometric}
\zerodisplayskips[1]
\cL_{photo}=\lambda\cL_{1}(\hat{\cI},\cI) +  (1-\lambda)SSIM(\hat{\cI},\cI)
\end{equation}
where $\hat{\cI}$ is the rendered image and $\cI$ is the ground truth image. Our EVPGS framework is generic and can be implemented with various GS-based methods as backbones. For each backbone, we also incorporate the specific strategies proposed in their respective papers.

\boldparagraph{Coarse Stage} We sample the augmented views by elevating the camera perspective at the training views for $\omega$ degrees. We formulate the coarse-stage loss function $\cL_{coarse}$ by summing the photometric loss in Eq.~\ref{eq:loss_photometric} 
and our AGR regularization loss terms in Eq.~\ref{eq:loss_reg_geometry} and \ref{eq:loss_reg_appearance}. During optimization, we update the GS model by taking gradients collectively on all loss terms in $\cL_{coarse}$, \ie, 
\begin{equation}\label{eq:loss_coarse}
\zerodisplayskips[1]
    \cL_{coarse} = \cL_{photo}^{train} + \lambda_{a}\cL_{a} + \lambda_{g}\cL_{g}
\end{equation}
where $\lambda_{a}$ and $\lambda_{g}$ are the loss balancing weights, $\cL_{photo}^{train}$ is the photometric loss applied to the training images. We apply $\cL_{a}$ and $\cL_{g}$ only to the augmented views.

\boldparagraph{Fine Stage}\label{sec:stage_fine}
At the fine stage, we use \textit{Enhanced View Priors} as pseudo-labels to fine-tune GS model on the augmented views. We employ the photometric loss $\cL_{photo}^{aug}$ with \textit{Enhanced View Prior} $\bar\cI$ as the supervision signal at the augmented views, \ie,
\begin{equation}\label{eq:loss_view_priors}
\zerodisplayskips[1]
    \cL_{photo}^{aug}=\lambda\cL_{1}(\hat{\cI},\bar{\cI}) +  (1-\lambda)SSIM(\hat{\cI},\bar{\cI})
\end{equation}
where $\hat{\cI}$ is the directly rendered images from the GS model.
We formulate the fine-stage loss $\cL_{fine}$ by incorporating the training view loss from Eq.~\ref{eq:loss_photometric} into Eq.~\ref{eq:loss_view_priors}, \ie,
\begin{equation}\label{eq:loss_fine}
\zerodisplayskips[1]
    \cL_{fine}=\cL_{photo}^{train}+\cL_{photo}^{aug}
\end{equation}

\section{Experiments}\label{sec:experiments}

\begin{table*}[]
    \centering
    \resizebox{0.95\linewidth}{!}{
    \begin{tabular}{@{}l@{\,\,}|ccc|ccc|ccc}
    & \multicolumn{3}{c|}{DTU} & \multicolumn{3}{c|}{Merchandise3D} & \multicolumn{3}{c}{Synthetic-NeRF} \\
    & PSNR $\uparrow$ & SSIM $\uparrow$ & LPIPS $\downarrow$ & PSNR $\uparrow$ & SSIM $\uparrow$ & LPIPS $\downarrow$ & PSNR $\uparrow$ & SSIM $\uparrow$ & LPIPS $\downarrow$\\ \hline
    3DGS\cite{kerbl20233dgaussiansplattingrealtime}&   25.188    &  0.8599 & 0.0865  & 23.946  & 0.9125  &  0.0618  & 24.479 & 0.8857  &  0.0804 
\\
Mip-Splatting\cite{yu2024mip-splatting} & 25.443 & 0.8668 & 0.0851 & 24.276  & 0.9139  & 0.0607  & 27.507 & 0.9134 &  \cellcolor{yellow}0.0517		
\\
2DGS\cite{huang20242d}&  25.589 & 0.8843 & 0.0762 & 22.181  & 0.9037  & 0.0746  & 27.058 & 0.9137 & 0.0573
\\
GOF\cite{Yu2024GOF}&  25.874 & 0.8633 & 0.0756 & 22.903  & 0.9169  & 0.0639  & 27.528 & 0.9170 & 0.0530
\\
RaDe-GS\cite{zhang2024radegsrasterizingdepthgaussian} & 25.778 & 0.8882 & 0.0753 & 23.559 & 0.9134  &  0.0645 & 27.531 & 0.9190 & 0.0530
\\
\hline
VEGS$^{*}$\cite{hwang2024vegsviewextrapolationurban}
&24.927 &
0.8674 &
0.0985 &
24.039  &
0.9115 &
0.0654  &
24.509 &
0.8935 &
0.0944 
\\
\hline
EVPGS(3DGS)
& 26.101 
& 0.8784
& 0.0773
& \cellcolor{orange}24.789 & 0.9204 & \cellcolor{orange}0.0563 & 27.270 &0.9173 &	0.0540
\\
EVPGS(MS)
& \cellcolor{yellow}26.107 
& 0.8777
& 0.0802
& \cellcolor{yellow}24.736 & \cellcolor{yellow}0.9207 & \cellcolor{orange}0.0563 & \cellcolor{orange}27.735 &	\cellcolor{yellow}0.9206 &	\cellcolor{orange}0.0507
\\
EVPGS(2DGS)
& 26.028 & \cellcolor{yellow}0.8936  & \cellcolor{yellow}0.0715 & 24.554 & 0.9194  &  \cellcolor{yellow}0.0574 & 27.266 & 0.9183 & 0.0565
\\
EVPGS(GOF)
& \cellcolor{orange}26.343 
& \cellcolor{orange}0.8959  
& \cellcolor{orange}0.0702 
& 23.565
& \cellcolor{orange}0.9239  
& 0.0577  
& \cellcolor{yellow}27.666 
& \cellcolor{orange}0.9221 
& 0.0526
\\

EVPGS(RaDe-GS)
&\cellcolor{tablered}26.488 &
\cellcolor{tablered}0.8991 &
\cellcolor{tablered}0.0670 &
\cellcolor{tablered}25.136  &
\cellcolor{tablered}0.9267 &
\cellcolor{tablered}0.0496  &
\cellcolor{tablered}27.849 &
\cellcolor{tablered}0.9243 &
\cellcolor{tablered}  0.0498 
    \end{tabular}
    }
    \vspace{-0.1in}
    \caption{\textbf{Quantitative evaluation of our EVPGS framework.} We compare the EVPGS variants based on different backbones with the baselines on the DTU~\cite{jensen2014large}, Merchandise3D, and Synthetic-NeRF~\cite{mildenhall2020nerfrepresentingscenesneural} datasets. For each dataset, we report the average metrics across all scenes. Each EVPGS variant substantially improves performance over the corresponding baselines.
    }
    \label{tab:main_comparison}
    \vspace{-15pt}
\end{table*}

\begin{figure*}
    \centering
    \includegraphics[width=1.0\linewidth]{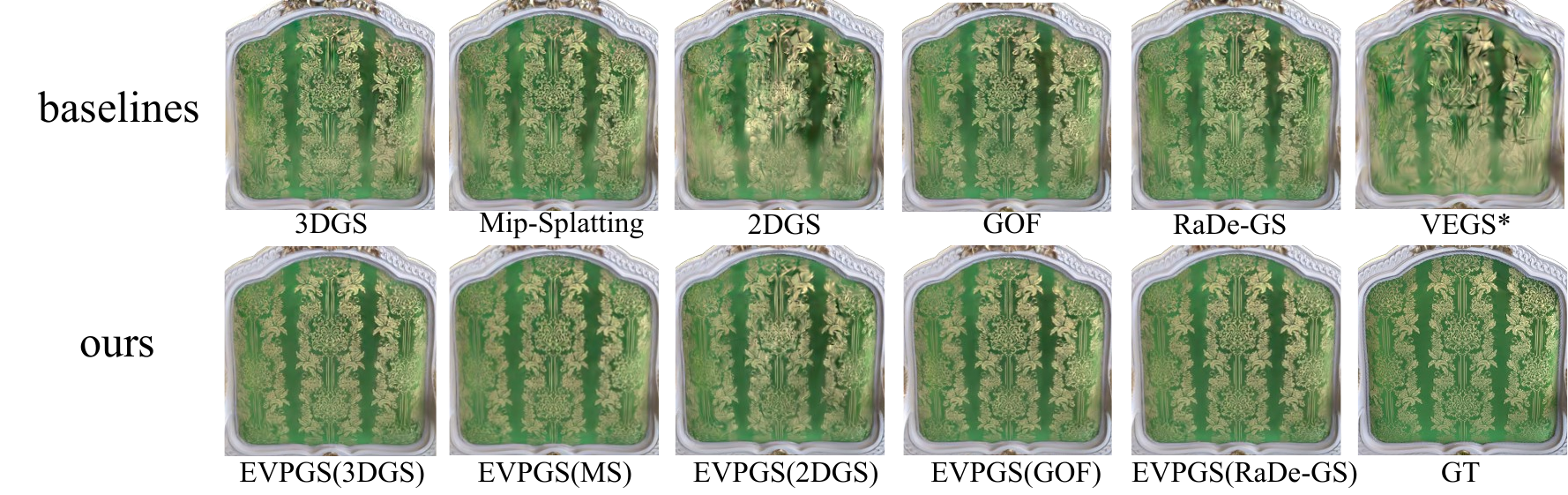}
    \caption{\textbf{Qualitative results on the synthetic dataset Synthetic-NeRF~\cite{mildenhall2020nerfrepresentingscenesneural}.} When using different backbones, our EVPGS produces synthesis closer to the ground truth than the baselines at extrapolated views.}
    \vspace{-15pt}
    \label{fig:syn_compasison}
\end{figure*}

\subsection{Experimental Setup}
\label{Sec:setup}

\boldparagraph{Datasets} We conduct experiments on three datasets with training/testing splits organized following the EVS scenario. These include the two public datasets DTU \cite{jensen2014large} and Synthetic-NeRF \cite{mildenhall2020nerfrepresentingscenesneural}, and our custom-captured real-world dataset called Merchandise3D which closely reflects the EVS scenario.
Merchandise3D includes seven common products in real life with a significant difference in view coverage between training views and testing views. 
More details of our dataset are in supplementary material.

\begin{enumerate}[label={\arabic*)}]
\item \textbf{DTU}\cite{jensen2014large}\textbf{:} We select 14 scanned scenes in our EVS task. 
Each scan contains either 49 or 64 views. 
For each scene, we use the 21 or 31 views with the smallest pitch angles for training and the remaining for testing.
\item \textbf{Merchandise3D:} For each object, we use 50 images captured from horizontal viewpoints as training views and the remaining 50 images from elevated viewpoints as testing views.
\item \textbf{Synthetic-NeRF}\cite{mildenhall2020nerfrepresentingscenesneural}\textbf{:} 
We sort the camera locations of the original training/testing split in ascending order along the z-axis to select the lowest 100 views as training views, leaving the remaining 200 views for testing.
\end{enumerate}

We quantify the view coverage disparity between the training and testing splits by calculating the pitch angle difference between each test view and its closest training view, averaged across all testing views for each dataset, as shown in Table~\ref{tab:angle_comparison}.

\begin{table}[]
    \centering
    \resizebox{1.0\linewidth}{!}{
    \begin{tabular}{|c|cllcll|}
\hline
\multirow{2}{*}{\textbf{Dataset}} & \multicolumn{6}{c|}{\textbf{Average Training/Testing Pitch Angle Difference (Degree)}}              \\ \cline{2-7}  & \multicolumn{3}{c|}{Original Split}  & \multicolumn{3}{c|}{Our EVS-specific split} \\ \hline
DTU~\cite{jensen2014large} & \multicolumn{3}{c|}{11.576} & \multicolumn{3}{c|}{\textbf{24.497}}                     \\ 
\hline
Synthetic-NeRF~\cite{mildenhall2020nerfrepresentingscenesneural} & \multicolumn{3}{c|}{8.463}       & \multicolumn{3}{c|}{\textbf{26.491}}                  \\ \hline
Merchandise3D  &\multicolumn{3}{c|}{-} &\multicolumn{3}{c|}{\textbf{39.580}} \\ \hline
\end{tabular}
    }
    \caption{
    \textbf{Average pitch angle difference between the original training/testing split and our EVS-specific split.} We measure the difference for the three datasets DTU, Synthetic-NeRF, and Merchandise3D. Our EVS splits exhibit a larger view disparity on each dataset, posing a more challenging scenario.}
    
    \label{tab:angle_comparison}
    \vspace{-18pt}
\end{table}

\boldparagraph{Baselines and metrics} 
We use six GS-based methods as baselines, \ie 3DGS~\cite{kerbl20233dgaussiansplattingrealtime}, Mip-Splatting~\cite{yu2024mip-splatting}, 2DGS~\cite{huang20242d}, GOF~\cite{Yu2024GOF}, RaDe-GS~\cite{zhang2024radegsrasterizingdepthgaussian},
and VEGS~\cite{hwang2024vegsviewextrapolationurban}. These baselines are trained only on the training views for each scene in each dataset.
To evaluate the flexibility of our EVPGS framework, 
we integrate it with 
each GS-based backbone except VEGS, as VEGS is specifically designed to tackle the EVS problem. Based on the results, we select RaDe-GS as the primary backbone for EVPGS, as it achieves the best performance in our experiments.
In the following figures and tables, we use \textit{EVPGS(Backbone)} to denote each backbone in the corresponding experiment.
Following the baseline methods, we adopt PSNR, SSIM \cite{wang2004image}, and LPIPS \cite{zhang2018unreasonable} as quantitative metrics.

\boldparagraph{Implementation Details} 
We first pre-train the backbone model for 20k iterations.
Then at the coarse stage, we employ the Stable-Diffusion-2.1 model \cite{rombach2022highresolutionimagesynthesislatent} trained on billion-scale real-world images to provide strong knowledge priors for our AGR strategy. We train for 1k iterations in this stage. 
At the fine stage, we set the reverse-reprojection error threshold $\xi=1$ and train for 9k iterations. We will explain more implementation details in supplementary.

\begin{figure*}
    \centering
    \includegraphics[width=1.0\linewidth]{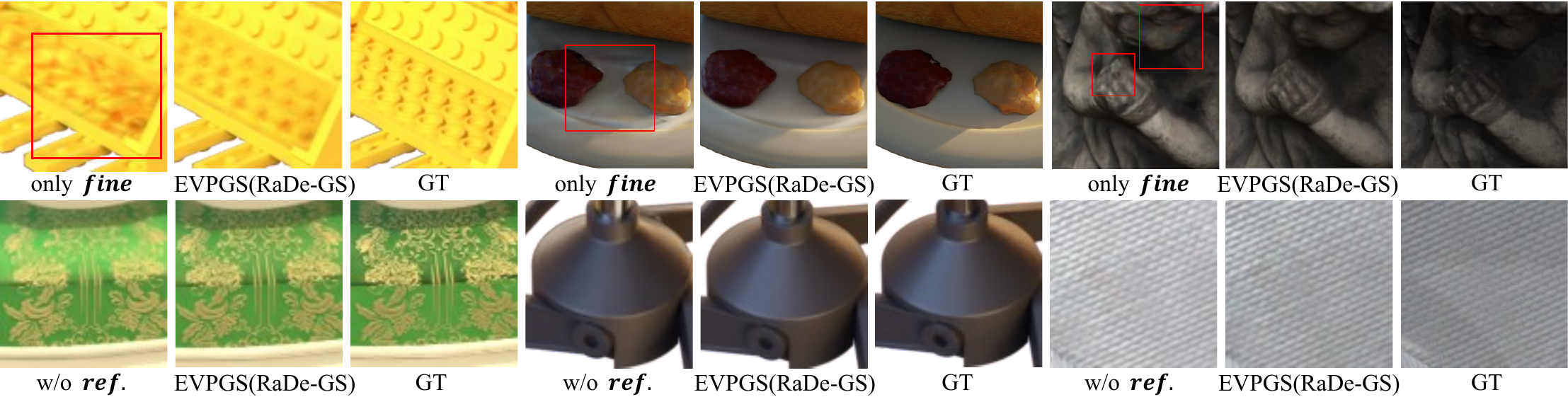}
    \vspace{-15pt}
    \caption{\textbf{Qualitative results from the ablation studies.} The first row demonstrates the effectiveness of the coarse stage in reducing prominent artifacts. The second row illustrates the impact of the view prior refinement strategy, which effectively combines view-dependent color information from the coarse stage while preserving fine-grained details.}
    \vspace{-15pt}
    \label{fig:ablation}
\end{figure*}

\subsection{Comparison to baselines}\label{sec:quantitative}
We compare our EVPGS framework with 3DGS \cite{kerbl20233dgaussiansplattingrealtime} and 
four
of its variants, \ie Mip-Splatting \cite{yu2024mip-splatting}, 2DGS \cite{huang20242d}, 
GOF~\cite{Yu2024GOF}
and RaDe-GS \cite{zhang2024radegsrasterizingdepthgaussian}. 
To further evaluate the generality of EVPGS for the EVS task, we adopt these
five GS methods as alternative backbones. 
Additionally, we compare EVPGS with VEGS~\cite{hwang2024vegsviewextrapolationurban}, a GS-based method specifically designed for the EVS problem in urban scenes. To adapt VEGS to our scenario, we remove the LiDAR initialization and object detection, referring to this variant as VEGS$^{*}$ in the following.

Table \ref{tab:main_comparison} shows quantitative results on DTU~\cite{jensen2014large}, Merchandise3D, and Synthetic-NeRF~\cite{mildenhall2020nerfrepresentingscenesneural}. 
EVPGS(RaDe-GS)
consistently outperforms the baseline methods across all metrics, including PSNR, SSIM, and LPIPS, on all three datasets. Integrating the
five baseline models into our EVPGS framework brings substantial performance improvements over the corresponding baselines.

Figure~\ref{fig:real_compasison} shows the qualitative results on the real-world datasets DTU and Merchandise3D. In comparison with the baselines, our EVPGS framework can synthesize extrapolated images with fewer and milder artifacts, while preserving high-frequency details (\eg text on the merchandise and tiles on the roof). These results highlight the ability of EVPGS to recover fine-grained details and produce realistic extrapolated views. Moreover, we show in Figure~\ref{fig:syn_compasison} our qualitative improvements on the synthetic dataset Synthetic-NeRF. The results on Synthetic-NeRF also demonstrate the efficacy of EVPGS. 

\subsection{Ablation Study}\label{sec:ablation}

\begin{table}[]
    \centering
    \resizebox{1.0\linewidth}{!}{
    \begin{tabular}{@{}l@{\,\,}|ccc|ccc}
    & \multicolumn{3}{c|}{DTU} &
    \multicolumn{3}{c}{Synthetic-NeRF} \\
    & \!PSNR $\uparrow$\! & \!SSIM $\uparrow$\! & \!LPIPS $\downarrow$ &  \!PSNR $\uparrow$\! & \!SSIM $\uparrow$\! & \!LPIPS $\downarrow$\!\\ 
    \hline
    $baseline$ & 25.778 & 0.8882 & 0.0753 & 27.531 & 0.9190 & 0.0530 \\
\hline
$only \: coarse$ & 26.219 & 0.8927 & 0.0699 & 27.683 & 0.9222 & 0.0507\\
$only \: fine$ & 26.165 & 0.8978 & 0.0703 & 27.708 & 0.9224 & 0.0523\\ 
$full \: w/o\: occ.$ & 26.027 & 0.8791 & 0.0714 & 27.606 & 0.9217 & 0.0519\\
$full \: w/o\: ref.$ & 26.155 & 0.8951 & 0.0714 & 27.457 & 0.9221 & 0.0514 \\
\hline
$full\: w/o\: mesh$ & 26.337 & 0.8956 & 0.0680 & 27.829 & 0.9237 & 0.0501 \\
\hline
$full$ & \textbf{26.488} & \textbf{0.8991} & \textbf{0.0670} & \textbf{27.849} & \textbf{0.9243} & \textbf{0.0498}  \!\!\!
    \end{tabular}
    }
    \vspace{-10pt}
    \caption{
    \textbf{Ablation study on DTU~\cite{jensen2014large} and Synthetic-NeRF~\cite{mildenhall2020nerfrepresentingscenesneural}.} We use RaDe-GS~\cite{zhang2024radegsrasterizingdepthgaussian} as the baseline.
    }
    \label{tab:main_ablation}
    \vspace{-15pt}
\end{table}

\boldparagraph{Ablation on both coarse and fine stages}
We investigate the impact of each proposed strategies on EVS performance.  
Following the results in Table~\ref{tab:main_ablation}, we use RaDe-GS~\cite{zhang2024radegsrasterizingdepthgaussian} as the backbone in our ablation study, since EVPGS(RaDe-GS) brings the best performance. We evaluate the effectiveness of both coarse and fine stages in our framework, as well as the two strategies of the fine stage, \ie occlusion-aware reprojection and view prior refinement. 
For the coarse and fine stages, we conduct two separate experiments where each experiment has only one of the stages applied. We denote these two simplified versions of EVPGS as $only \: coarse$ and $only \: fine$, respectively. For the two strategies of the fine stage, we conduct another two experiments by removing either strategy from the complete EVPGS framework, resulting in two more simplified versions of EVPGS which we denote as $full \: w/o \: occ.$ and $full \: w/o \: ref.$, respectively. As shown in Table~\ref{tab:main_ablation}, all of the proposed components contribute significantly to EVPGS.

We also provide qualitative comparison in Figure~\ref{fig:ablation} to demonstrate the effectiveness of both the coarse stage and the view prior refinement
strategy.
The coarse stage removes severe artifacts, therefore benefiting the subsequent fine stage. View prior refinement helps preserve 
more view-specific visual information
in the final rendered images.

\boldparagraph{Ablation on initial reconstructed mesh}
We also investigate how the usage of the reconstructed mesh benefits both the coarse and fine stages.
In our pipeline, we reconstruct an initial mesh after pre-training the GS backbone to enhance depth map quality, addressing the cases where GS-rendered depth maps are overall relatively accurate but incomplete in some augmented views. 
To account for scenarios 
where the initial reconstructed mesh is inaccurate or unavailable (e.g., complex objects or outdoor scenes), 
we conduct experiments on a variant of our EVPGS which does not rely on the initial mesh.
Specifically, we remove the geometry regularization in AGR module (Sec.~\ref{Sec:AGR}) and rely 
the GS-rendered depth map instead of the mesh-rendered depth map for occlusion detection (Sec.~\ref{Sec:RRP}). We denote this version of EVPGS as $full \: w/o \: mesh$. As shown in Table~\ref{tab:main_ablation}, $full \: w/o \: mesh$ has moderate performance drop compared to the full EVPGS, while still outperforming the baseline as well as the other simplified variants. Incorporating the reconstructed mesh in both the coarse and fine stages contribute to improved performance.
\section{Limitations}\label{sec:limitations}
There are two limitations to address in future work.
While we handle view-dependent color at augmented views using the OARR strategy (Sec.~\ref{sec:fusion}), our method does not account for
the part of view dependence caused by complex lighting effects as we do not explicitly model lighting. 
Despite the high-fidelity renderings at extrapolated views produced
without considering lighting, there remains a noticeable color discrepancy compared to the ground truth, which can affect the overall visual experience.  
The other limitation of our method is its reliance on the quality of the reconstructed mesh. 
While our ablation study in Sec.~\ref{sec:ablation} shows that EVPGS still performs competitively without an initial mesh, we aim to explore mesh-free approaches in future work 
that are better applicable to scenarios where meshes are inaccessible.
\section{Conclusion}\label{sec:conclusion}
In this work, we proposed the view augmentation framework called EVPGS based on Gaussian Splatting (GS) to tackle the challenging Extrapolated View Synthesis (EVS) problem. EVPGS generates enhanced view priors to fine-tune GS models through a coarse-to-fine process. At the coarse stage, we proposed the Appearance and Geometry Regularization (AGR) strategy to effectively reduce severe rendering artifacts. At the fine stage, we proposed the Occlusion-Aware Reprojection and Refinement (OARR) strategy to generate enhanced view priors, which serves as strong appearance guidance for synthesis at extrapolated views. 
To comprehensively evaluate our EVPGS framework, we conducted experiments on two public real-world and synthetic datasets, \ie DTU, Synthetic-NeRF, and our self-collected real-world dataset Merchandise3D. Both quantitative and qualitative results on all three datasets demonstrate that EVPGS significantly improves the synthesis quality at extrapolated views compared to the baseline methods. Our EVPGS framework is generic and achieves 
consistent improvements across different backbones.

{
    \small
    \bibliographystyle{ieeenat_fullname}
    \bibliography{main}

\begin{thebibliography}{50}
\providecommand{\natexlab}[1]{#1}
\providecommand{\url}[1]{\texttt{#1}}
\expandafter\ifx\csname urlstyle\endcsname\relax
  \providecommand{\doi}[1]{doi: #1}\else
  \providecommand{\doi}{doi: \begingroup \urlstyle{rm}\Url}\fi

\bibitem[Barron et~al.(2021)Barron, Mildenhall, Tancik, Hedman, Martin-Brualla, and Srinivasan]{barron2021mip}
Jonathan~T Barron, Ben Mildenhall, Matthew Tancik, Peter Hedman, Ricardo Martin-Brualla, and Pratul~P Srinivasan.
\newblock Mip-nerf: A multiscale representation for anti-aliasing neural radiance fields.
\newblock In \emph{Proceedings of the IEEE/CVF international conference on computer vision}, pages 5855--5864, 2021.

\bibitem[Barron et~al.(2022)Barron, Mildenhall, Verbin, Srinivasan, and Hedman]{barron2022mip}
Jonathan~T Barron, Ben Mildenhall, Dor Verbin, Pratul~P Srinivasan, and Peter Hedman.
\newblock Mip-nerf 360: Unbounded anti-aliased neural radiance fields.
\newblock In \emph{Proceedings of the IEEE/CVF conference on computer vision and pattern recognition}, pages 5470--5479, 2022.

\bibitem[Charatan et~al.(2024)Charatan, Li, Tagliasacchi, and Sitzmann]{charatan2024pixelsplat3dgaussiansplats}
David Charatan, Sizhe Li, Andrea Tagliasacchi, and Vincent Sitzmann.
\newblock pixelsplat: 3d gaussian splats from image pairs for scalable generalizable 3d reconstruction, 2024.

\bibitem[Chen et~al.(2022)Chen, Xu, Geiger, Yu, and Su]{chen2022tensorftensorialradiancefields}
Anpei Chen, Zexiang Xu, Andreas Geiger, Jingyi Yu, and Hao Su.
\newblock Tensorf: Tensorial radiance fields, 2022.

\bibitem[Chen et~al.(2024)Chen, Xu, Zheng, Zhuang, Pollefeys, Geiger, Cham, and Cai]{chen2024mvsplat}
Yuedong Chen, Haofei Xu, Chuanxia Zheng, Bohan Zhuang, Marc Pollefeys, Andreas Geiger, Tat-Jen Cham, and Jianfei Cai.
\newblock Mvsplat: Efficient 3d gaussian splatting from sparse multi-view images.
\newblock In \emph{ECCV}, 2024.

\bibitem[Choi et~al.(2019)Choi, Gallo, Troccoli, Kim, and Kautz]{choi2019extreme}
Inchang Choi, Orazio Gallo, Alejandro Troccoli, Min~H Kim, and Jan Kautz.
\newblock Extreme view synthesis.
\newblock In \emph{Proceedings of the IEEE/CVF International Conference on Computer Vision}, pages 7781--7790, 2019.

\bibitem[Fridovich-Keil et~al.(2022)Fridovich-Keil, Yu, Tancik, Chen, Recht, and Kanazawa]{yu2022plenoxels}
Sara Fridovich-Keil, Alex Yu, Matthew Tancik, Qinhong Chen, Benjamin Recht, and Angjoo Kanazawa.
\newblock Plenoxels: Radiance fields without neural networks.
\newblock In \emph{CVPR}, 2022.

\bibitem[Fridovich-Keil et~al.(2023)Fridovich-Keil, Meanti, Warburg, Recht, and Kanazawa]{fridovichkeil2023kplanesexplicitradiancefields}
Sara Fridovich-Keil, Giacomo Meanti, Frederik Warburg, Benjamin Recht, and Angjoo Kanazawa.
\newblock K-planes: Explicit radiance fields in space, time, and appearance, 2023.

\bibitem[Gonzalez and Woods(2008)]{gonzalez2008}
Rafael~C. Gonzalez and Richard~E. Woods.
\newblock \emph{Digital Image Processing (3rd Edition)}.
\newblock Prentice Hall, 2008.

\bibitem[Gu{\'e}don and Lepetit(2024)]{guedon2024sugar}
Antoine Gu{\'e}don and Vincent Lepetit.
\newblock Sugar: Surface-aligned gaussian splatting for efficient 3d mesh reconstruction and high-quality mesh rendering.
\newblock In \emph{Proceedings of the IEEE/CVF Conference on Computer Vision and Pattern Recognition}, pages 5354--5363, 2024.

\bibitem[Hedman et~al.(2021)Hedman, Srinivasan, Mildenhall, Barron, and Debevec]{hedman2021snerg}
Peter Hedman, Pratul~P. Srinivasan, Ben Mildenhall, Jonathan~T. Barron, and Paul Debevec.
\newblock Baking neural radiance fields for real-time view synthesis.
\newblock In \emph{ICCV}, 2021.

\bibitem[Ho et~al.(2020)Ho, Jain, and Abbeel]{ho2020denoisingdiffusionprobabilisticmodels}
Jonathan Ho, Ajay Jain, and Pieter Abbeel.
\newblock Denoising diffusion probabilistic models, 2020.

\bibitem[Huang et~al.(2024)Huang, Yu, Chen, Geiger, and Gao]{huang20242d}
Binbin Huang, Zehao Yu, Anpei Chen, Andreas Geiger, and Shenghua Gao.
\newblock 2d gaussian splatting for geometrically accurate radiance fields.
\newblock In \emph{ACM SIGGRAPH 2024 Conference Papers}, pages 1--11, 2024.

\bibitem[Hwang et~al.(2024)Hwang, Kim, Kang, Kang, and Choo]{hwang2024vegsviewextrapolationurban}
Sungwon Hwang, Min-Jung Kim, Taewoong Kang, Jayeon Kang, and Jaegul Choo.
\newblock Vegs: View extrapolation of urban scenes in 3d gaussian splatting using learned priors, 2024.

\bibitem[Jensen et~al.(2014)Jensen, Dahl, Vogiatzis, Tola, and Aan{\ae}s]{jensen2014large}
Rasmus Jensen, Anders Dahl, George Vogiatzis, Engin Tola, and Henrik Aan{\ae}s.
\newblock Large scale multi-view stereopsis evaluation.
\newblock In \emph{Proceedings of the IEEE conference on computer vision and pattern recognition}, pages 406--413, 2014.

\bibitem[Kerbl et~al.(2023)Kerbl, Kopanas, Leimkühler, and Drettakis]{kerbl20233dgaussiansplattingrealtime}
Bernhard Kerbl, Georgios Kopanas, Thomas Leimkühler, and George Drettakis.
\newblock 3d gaussian splatting for real-time radiance field rendering, 2023.

\bibitem[Kerbl et~al.(2024)Kerbl, Meuleman, Kopanas, Wimmer, Lanvin, and Drettakis]{kerbl2024hierarchical}
Bernhard Kerbl, Andreas Meuleman, Georgios Kopanas, Michael Wimmer, Alexandre Lanvin, and George Drettakis.
\newblock A hierarchical 3d gaussian representation for real-time rendering of very large datasets.
\newblock \emph{ACM Transactions on Graphics (TOG)}, 43\penalty0 (4):\penalty0 1--15, 2024.

\bibitem[Kirillov et~al.(2023)Kirillov, Mintun, Ravi, Mao, Rolland, Gustafson, Xiao, Whitehead, Berg, Lo, Dollar, and Girshick]{kirillov2023sam}
Alexander Kirillov, Eric Mintun, Nikhila Ravi, Hanzi Mao, Chloe Rolland, Laura Gustafson, Tete Xiao, Spencer Whitehead, Alexander~C. Berg, Wan-Yen Lo, Piotr Dollar, and Ross Girshick.
\newblock Segment anything.
\newblock In \emph{Proceedings of the IEEE/CVF International Conference on Computer Vision (ICCV)}, pages 4015--4026, 2023.

\bibitem[Li et~al.(2024)Li, Zhang, Bai, Zheng, Ning, Zhou, and Gu]{li2024dngaussian}
Jiahe Li, Jiawei Zhang, Xiao Bai, Jin Zheng, Xin Ning, Jun Zhou, and Lin Gu.
\newblock Dngaussian: Optimizing sparse-view 3d gaussian radiance fields with global-local depth normalization.
\newblock In \emph{Proceedings of the IEEE/CVF Conference on Computer Vision and Pattern Recognition}, pages 20775--20785, 2024.

\bibitem[Liu et~al.(2024)Liu, Chen, Kao, Tai, and Tang]{liu2024deceptive}
Xinhang Liu, Jiaben Chen, Shiu-Hong Kao, Yu-Wing Tai, and Chi-Keung Tang.
\newblock Deceptive-nerf/3dgs: Diffusion-generated pseudo-observations for high-quality sparse-view reconstruction.
\newblock In \emph{ECCV}, 2024.

\bibitem[Mallick et~al.(2024)Mallick, Goel, Kerbl, Carrasco, Steinberger, and De~La~Torre]{mallick2024taming}
Saswat~Subhajyoti Mallick, Rahul Goel, Bernhard Kerbl, Francisco~Vicente Carrasco, Markus Steinberger, and Fernando De~La~Torre.
\newblock Taming 3dgs: High-quality radiance fields with limited resources.
\newblock \emph{arXiv preprint arXiv:2406.15643}, 2024.

\bibitem[Mildenhall et~al.(2020)Mildenhall, Srinivasan, Tancik, Barron, Ramamoorthi, and Ng]{mildenhall2020nerfrepresentingscenesneural}
Ben Mildenhall, Pratul~P. Srinivasan, Matthew Tancik, Jonathan~T. Barron, Ravi Ramamoorthi, and Ren Ng.
\newblock Nerf: Representing scenes as neural radiance fields for view synthesis, 2020.

\bibitem[M\"uller et~al.(2022)M\"uller, Evans, Schied, and Keller]{mueller2022instant-ngp}
Thomas M\"uller, Alex Evans, Christoph Schied, and Alexander Keller.
\newblock Instant neural graphics primitives with a multiresolution hash encoding.
\newblock \emph{ACM Trans. Graph.}, 41\penalty0 (4):\penalty0 102:1--102:15, 2022.

\bibitem[Niemeyer et~al.(2021)Niemeyer, Barron, Mildenhall, Sajjadi, Geiger, and Radwan]{niemeyer2021regnerfregularizingneuralradiance}
Michael Niemeyer, Jonathan~T. Barron, Ben Mildenhall, Mehdi S.~M. Sajjadi, Andreas Geiger, and Noha Radwan.
\newblock Regnerf: Regularizing neural radiance fields for view synthesis from sparse inputs, 2021.

\bibitem[Paliwal et~al.(2025)Paliwal, Ye, Xiong, Kotovenko, Ranjan, Chandra, and Kalantari]{paliwal2025coherentgs}
Avinash Paliwal, Wei Ye, Jinhui Xiong, Dmytro Kotovenko, Rakesh Ranjan, Vikas Chandra, and Nima~Khademi Kalantari.
\newblock Coherentgs: Sparse novel view synthesis with coherent 3d gaussians.
\newblock In \emph{European Conference on Computer Vision}, pages 19--37. Springer, 2025.

\bibitem[Radl et~al.(2024)Radl, Steiner, Parger, Weinrauch, Kerbl, and Steinberger]{radl2024stopthepop}
Lukas Radl, Michael Steiner, Mathias Parger, Alexander Weinrauch, Bernhard Kerbl, and Markus Steinberger.
\newblock Stopthepop: Sorted gaussian splatting for view-consistent real-time rendering.
\newblock \emph{ACM Transactions on Graphics (TOG)}, 43\penalty0 (4):\penalty0 1--17, 2024.

\bibitem[Rombach et~al.(2022)Rombach, Blattmann, Lorenz, Esser, and Ommer]{rombach2022highresolutionimagesynthesislatent}
Robin Rombach, Andreas Blattmann, Dominik Lorenz, Patrick Esser, and Björn Ommer.
\newblock High-resolution image synthesis with latent diffusion models, 2022.

\bibitem[Sarlin et~al.(2019)Sarlin, Cadena, Siegwart, and Dymczyk]{sarlin2019coarse}
Paul-Edouard Sarlin, Cesar Cadena, Roland Siegwart, and Marcin Dymczyk.
\newblock From coarse to fine: Robust hierarchical localization at large scale.
\newblock In \emph{CVPR}, 2019.

\bibitem[Schonberger and Frahm(2016)]{schonberger2016structure}
Johannes~L Schonberger and Jan-Michael Frahm.
\newblock Structure-from-motion revisited.
\newblock In \emph{Proceedings of the IEEE conference on computer vision and pattern recognition}, pages 4104--4113, 2016.

\bibitem[Vincent(2011)]{vincent2011connection}
Pascal Vincent.
\newblock A connection between score matching and denoising autoencoders.
\newblock \emph{Neural computation}, 23\penalty0 (7):\penalty0 1661--1674, 2011.

\bibitem[Wang et~al.(2023{\natexlab{a}})Wang, Chen, Loy, and Liu]{wang2023sparsenerfdistillingdepthranking}
Guangcong Wang, Zhaoxi Chen, Chen~Change Loy, and Ziwei Liu.
\newblock Sparsenerf: Distilling depth ranking for few-shot novel view synthesis, 2023{\natexlab{a}}.

\bibitem[Wang et~al.(2021)Wang, Liu, Liu, Theobalt, Komura, and Wang]{wang2021neus}
Peng Wang, Lingjie Liu, Yuan Liu, Christian Theobalt, Taku Komura, and Wenping Wang.
\newblock Neus: Learning neural implicit surfaces by volume rendering for multi-view reconstruction.
\newblock In \emph{NIPS}, 2021.

\bibitem[Wang et~al.(2023{\natexlab{b}})Wang, Han, Habermann, Daniilidis, Theobalt, and Liu]{wang2023neus2}
Yiming Wang, Qin Han, Marc Habermann, Kostas Daniilidis, Christian Theobalt, and Lingjie Liu.
\newblock Neus2: Fast learning of neural implicit surfaces for multi-view reconstruction.
\newblock In \emph{Proceedings of the IEEE/CVF International Conference on Computer Vision (ICCV)}, 2023{\natexlab{b}}.

\bibitem[Wang et~al.(2024)Wang, Huang, Chen, and Lee]{wang2024freesplat}
Yunsong Wang, Tianxin Huang, Hanlin Chen, and Gim~Hee Lee.
\newblock Freesplat: Generalizable 3d gaussian splatting towards free-view synthesis of indoor scenes.
\newblock \emph{arXiv preprint arXiv:2405.17958}, 2024.

\bibitem[Wang et~al.(2004)Wang, Bovik, Sheikh, and Simoncelli]{wang2004image}
Zhou Wang, Alan~C Bovik, Hamid~R Sheikh, and Eero~P Simoncelli.
\newblock Image quality assessment: from error visibility to structural similarity.
\newblock \emph{IEEE transactions on image processing}, 13\penalty0 (4):\penalty0 600--612, 2004.

\bibitem[Wu et~al.(2023)Wu, Wang, Pan, Xu, Theobalt, Liu, and Lin]{wu2023voxurf}
Tong Wu, Jiaqi Wang, Xingang Pan, Xudong Xu, Christian Theobalt, Ziwei Liu, and Dahua Lin.
\newblock Voxurf: Voxel-based efficient and accurate neural surface reconstruction.
\newblock In \emph{International Conference on Learning Representations (ICLR)}, 2023.

\bibitem[Wynn and Turmukhambetov(2023)]{wynn2023diffusionerfregularizingneuralradiance}
Jamie Wynn and Daniyar Turmukhambetov.
\newblock Diffusionerf: Regularizing neural radiance fields with denoising diffusion models, 2023.

\bibitem[Xu et~al.(2024)Xu, Gao, Shen, Peng, Jiao, and Wang]{xu2024mvpgsexcavatingmultiviewpriors}
Wangze Xu, Huachen Gao, Shihe Shen, Rui Peng, Jianbo Jiao, and Ronggang Wang.
\newblock Mvpgs: Excavating multi-view priors for gaussian splatting from sparse input views, 2024.

\bibitem[Yang et~al.(2023{\natexlab{a}})Yang, Li, Zhou, Yuan, Liu, Yang, Qiu, and Shen]{yang2023nerfvs}
Chen Yang, Peihao Li, Zanwei Zhou, Shanxin Yuan, Bingbing Liu, Xiaokang Yang, Weichao Qiu, and Wei Shen.
\newblock Nerfvs: Neural radiance fields for free view synthesis via geometry scaffolds.
\newblock In \emph{Proceedings of the IEEE/CVF Conference on Computer Vision and Pattern Recognition}, pages 16549--16558, 2023{\natexlab{a}}.

\bibitem[Yang et~al.(2023{\natexlab{b}})Yang, Pavone, and Wang]{yang2023freenerfimprovingfewshotneural}
Jiawei Yang, Marco Pavone, and Yue Wang.
\newblock Freenerf: Improving few-shot neural rendering with free frequency regularization, 2023{\natexlab{b}}.

\bibitem[Yariv et~al.(2021)Yariv, Gu, Kasten, and Lipman]{yariv2021volume}
Lior Yariv, Jiatao Gu, Yoni Kasten, and Yaron Lipman.
\newblock Volume rendering of neural implicit surfaces.
\newblock \emph{Advances in Neural Information Processing Systems}, 34:\penalty0 4805--4815, 2021.

\bibitem[Yu et~al.(2021)Yu, Li, Tancik, Li, Ng, and Kanazawa]{yu2021plenoctrees}
Alex Yu, Ruilong Li, Matthew Tancik, Hao Li, Ren Ng, and Angjoo Kanazawa.
\newblock {PlenOctrees} for real-time rendering of neural radiance fields.
\newblock In \emph{ICCV}, 2021.

\bibitem[Yu et~al.(2024{\natexlab{a}})Yu, Long, and Tan]{yu2024lmgaussianboostsparseview3d}
Hanyang Yu, Xiaoxiao Long, and Ping Tan.
\newblock Lm-gaussian: Boost sparse-view 3d gaussian splatting with large model priors, 2024{\natexlab{a}}.

\bibitem[Yu et~al.(2024{\natexlab{b}})Yu, Chen, Huang, Sattler, and Geiger]{yu2024mip-splatting}
Zehao Yu, Anpei Chen, Binbin Huang, Torsten Sattler, and Andreas Geiger.
\newblock Mip-splatting: Alias-free 3d gaussian splatting.
\newblock In \emph{Proceedings of the IEEE/CVF Conference on Computer Vision and Pattern Recognition (CVPR)}, pages 19447--19456, 2024{\natexlab{b}}.

\bibitem[Yu et~al.(2024{\natexlab{c}})Yu, Sattler, and Geiger]{Yu2024GOF}
Zehao Yu, Torsten Sattler, and Andreas Geiger.
\newblock Gaussian opacity fields: Efficient adaptive surface reconstruction in unbounded scenes.
\newblock \emph{ACM Transactions on Graphics}, 2024{\natexlab{c}}.

\bibitem[Zhang et~al.(2024)Zhang, Fang, Shrestha, Liang, Long, and Tan]{zhang2024radegsrasterizingdepthgaussian}
Baowen Zhang, Chuan Fang, Rakesh Shrestha, Yixun Liang, Xiaoxiao Long, and Ping Tan.
\newblock Rade-gs: Rasterizing depth in gaussian splatting, 2024.

\bibitem[Zhang et~al.(2022)Zhang, Zhang, Fu, Zhou, Cai, Huang, Jia, Zhao, and Tang]{zhang2022raypriors}
Jian Zhang, Yuanqing Zhang, Huan Fu, Xiaowei Zhou, Bowen Cai, Jinchi Huang, Rongfei Jia, Binqing Zhao, and Xing Tang.
\newblock Ray priors through reprojection: Improving neural radiance fields for novel view extrapolation.
\newblock In \emph{CVPR}, pages 18376--18386, 2022.

\bibitem[Zhang et~al.(2018)Zhang, Isola, Efros, Shechtman, and Wang]{zhang2018unreasonable}
Richard Zhang, Phillip Isola, Alexei~A Efros, Eli Shechtman, and Oliver Wang.
\newblock The unreasonable effectiveness of deep features as a perceptual metric.
\newblock In \emph{Proceedings of the IEEE conference on computer vision and pattern recognition}, pages 586--595, 2018.

\bibitem[Zhu et~al.(2025)Zhu, Fan, Jiang, and Wang]{zhu2025fsgs}
Zehao Zhu, Zhiwen Fan, Yifan Jiang, and Zhangyang Wang.
\newblock Fsgs: Real-time few-shot view synthesis using gaussian splatting.
\newblock In \emph{European Conference on Computer Vision}, pages 145--163. Springer, 2025.

\bibitem[Zwicker et~al.(2001)Zwicker, Pfister, Van~Baar, and Gross]{zwicker2001ewa}
Matthias Zwicker, Hanspeter Pfister, Jeroen Van~Baar, and Markus Gross.
\newblock Ewa volume splatting.
\newblock In \emph{Visualization, 2001. VIS 01. Proceedings}, pages 29--538. IEEE, 2001.

\end{thebibliography}
}

\setcounter{section}{0}
\renewcommand\thesection{\Alph{section}}
\setcounter{table}{0}
\renewcommand\thetable{\Alph{table}}
\setcounter{figure}{0}
\renewcommand\thefigure{\Alph{figure}}
\setcounter{equation}{0}

\clearpage
\setcounter{page}{1}
\maketitlesupplementary

In this \textbf{supplementary document}, we first provide a detailed overview of our Merchandise3D dataset, including its configuration and capture process, in Sec.\ref{sec:merchandise3d}. Next, we elaborate on further experimental details, including dataset specifics and training implementation, in Sec.\ref{sec:exp_details}. Subsequently, we present extended results for EVPGS, covering more quantitative and qualitative comparison experiments in Sec.\ref{sec:more_results}. Finally, we provide an example for the real-world application of EVPGS in Sec.\ref{sec:application}.

\section{Our Merchandise3D Dataset}\label{sec:merchandise3d}

\begin{figure}
    \centering
    \includegraphics[width=0.9\linewidth]{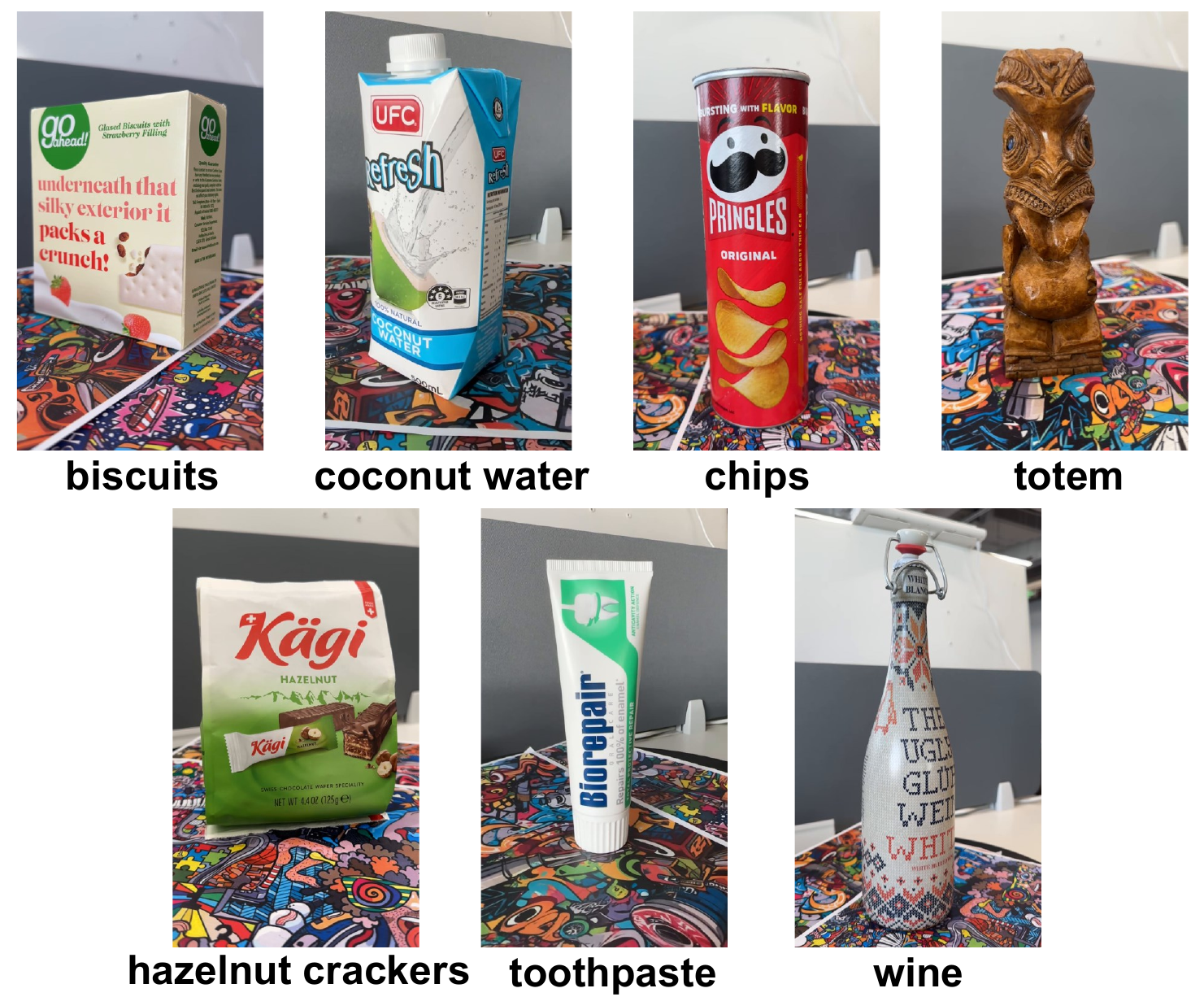}
    \caption{\textbf{Our Merchandise3D dataset}, which comprises real-world merchandise objects with diverse structures and textures. To boost the accuracy of HLOC~\cite{sarlin2019coarse}, we place graffiti sheets under each object during capturing as explained in Section~\ref{sec:merchandise3d}. We focus on EVS for objects in this work, and the graffiti is not a part of object to be reconstructed.}
    \label{fig:merchandise3d}
\end{figure}

To adequately represent both real-world applications and the challenges posed by EVS, our Merchandise3D dataset comprises real-world merchandise objects with diverse structures and textures. The dataset has 7 objects instances, each represented in the form of 100 camera views, where each view is a triplet that contains an RGB image, an object mask, and the corresponding camera parameters. 
As shown in Figure~\ref{fig:supp_evs} left, we select 50 views captured from nearly horizontal angles as the training set and another 50 views, elevated by approximately 40 degrees, as the testing set.
We instantiate each of these objects in Figure~\ref{fig:merchandise3d}. Each object is captured originally in the format of a video and we release these object videos as well.
The resolution of videos in our dataset is $1280\times720$.
We describe our video capturing and processing pipeline as follows.

When capturing video for each object, we place the object on a motorized rotating turntable and keep recording until the object is rotated for two rounds, which comprise one round from a horizontal viewpoint followed by another round from an elevated viewpoint looking downwards at approximately 40 degrees. We slowly lift the camera after the first round and keep the camera fixed otherwise.

When processing each video, we  sample 100 frames on average and solve the camera parameters for these frames using HLOC~\cite{sarlin2019coarse}. To boost the accuracy of HLOC, we introduce additional textures to the scene by placing graffiti sheets under the object when capturing the video. Each graffiti we select has non-repeating local patterns across the graffiti as to avoid confusing HLOC. When using multiple graffiti sheets, we use a different graffiti on each sheet for the same reason. Since we focus the view synthesis task on objects, we also provide the object mask of each frame obtained using the off-the-shelf Segment Anything Model (SAM)~\cite{kirillov2023sam}, while we treat the graffiti as background.

\begin{figure*}
    \centering
    \includegraphics[width=1.0\linewidth]{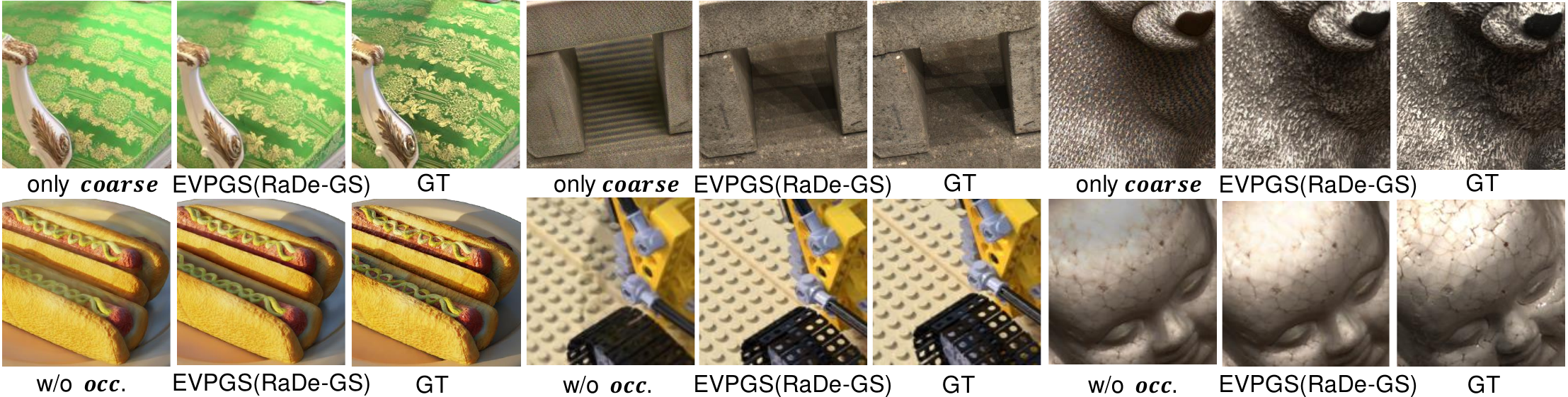}
    \caption{\textbf{More ablation study results}. Complementary to Figure~\co{6} in the main paper, we present additional qualitative ablation study results for the other two simplified version of EVPGS, \ie \textit{only coarse} and \textit{full w/o occ.} \textbf{Top row:} In comparison to the simplified EVPGS with only the coarse stage, the full EVPGS can better reconstruct fine details. \textbf{Bottom row:} In comparison to the simplified EVPGS without the occlusion-aware reprojection strategy, the full EVPGS produces renderings clear of corruptions caused by occlusion.
    }
    \label{fig:supp_ablation}
\end{figure*}

\section{Experiments Details}
\label{sec:exp_details}

\subsection{Datasets}

\boldparagraph{Details} 
For the two public datasets, DTU~\cite{jensen2014large} and Synthetic-NeRF~\cite{mildenhall2020nerfrepresentingscenesneural}, we utilize the ground truth masks provided by the datasets to segment objects from the background for the EVS task.
For the DTU~\cite{jensen2014large} dataset, we follow \cite{huang20242d, zhang2024radegsrasterizingdepthgaussian, Yu2024GOF} to evaluate the proposed method on 14 selected scenes. The selected scan IDs are 24, 37, 40, 55, 63, 65, 83, 97, 105, 106, 110, 114, 118, and 122. Scan 69 was excluded from our experiments as it lacks the ground truth masks necessary for segmenting the object from the background. To improve experimental efficiency, the image resolution is resized to $777\times 851$. For the Synthetic-NeRF~\cite{mildenhall2020nerfrepresentingscenesneural} dataset, all eight synthetic objects are included in our experiments. The training images keep the original resolution of $800\times 800$. For the Merchandise3D dataset, please refer to Sec.~\ref{sec:merchandise3d}

\begin{figure*}
    \centering
    \includegraphics[width=1.0\linewidth]{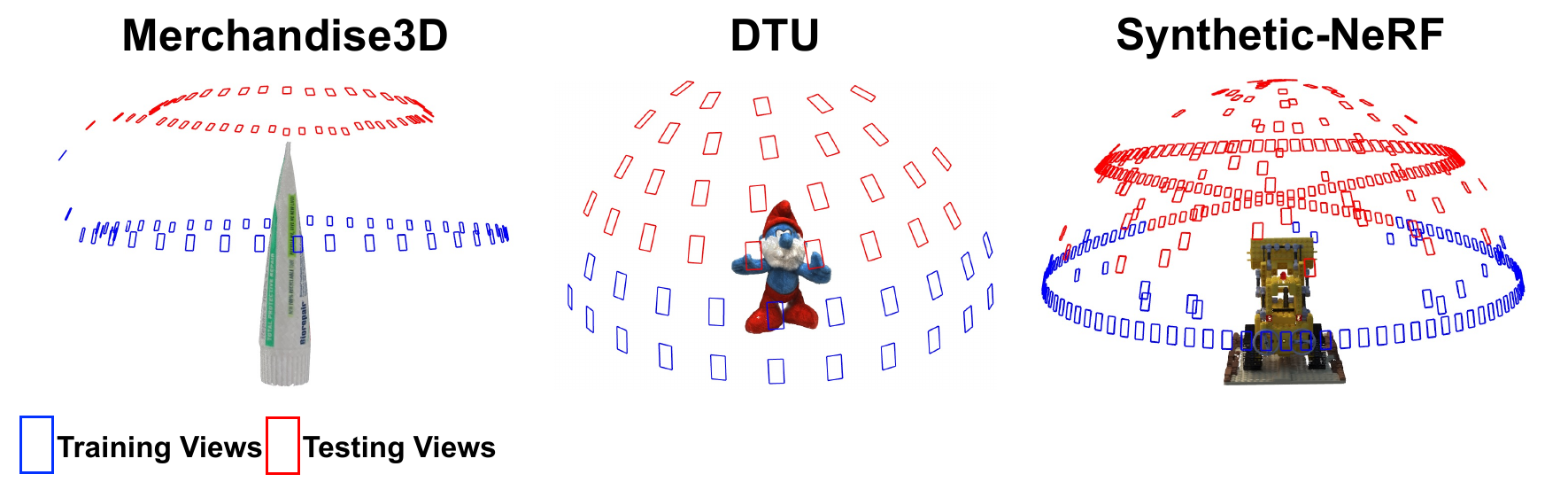}
    \caption{\textbf{Visualization of the substantial view coverage disparity between the training and testing splits in our EVS scenario.} The quantitative average angle differences are provided in Table~\co{2} of the main paper.}
    \vspace{-15pt}
    \label{fig:supp_evs}
\end{figure*}

\boldparagraph{Visualization} As shown in Table \co{2} of the main paper, each dataset organized following the EVS scenario exhibits drastic view coverage difference between the training and testing splits. We visualize some examples from each dataset in Figure~\ref{fig:supp_evs} to showcase the difficulty of the EVS problem.

\subsection{Training}

\boldparagraph{Coarse Stage} As mentioned in Sec.~\co{4} in the main paper, we use 3DGS~\cite{kerbl20233dgaussiansplattingrealtime}, Mip-Splatting~\cite{yu2024mip-splatting}, 2DGS~\cite{huang20242d}, GOF~\cite{Yu2024GOF} and RaDe-GS~\cite{zhang2024radegsrasterizingdepthgaussian} as alternative backbones for our EVPGS framework. When pre-training these GS-based models, we adhere to the training configurations specified in their respective original papers. For our appearance regularization (Sec.~\co{3.2} of the main paper), we utilize the Stable-Diffusion-2.1 model directly, without any additional fine-tuning on our dataset while using empty text prompts. 
For our geometry regularization (Sec.~\co{3.2} of the main paper), we use the \textit{pytorch3d} toolbox to rasterize the depth map from reconstructed mesh.
We set $\lambda_{a}=1e-7$ and $\lambda_{g}=1e-1$ in Eq.~\co{8} of the main paper. 

\boldparagraph{Fine stage} 
For our View Prior Refinement strategy (Section~\co{3.4} of the main paper), we select the parameters that bring the best performance for EVPGS.
We select $w_h=0.8$ and $w_l=0.5$.

\section{More Results}
\label{sec:more_results}

\subsection{Computational Cost}

We compute the training time of RaDe-GS~\cite{zhang2024radegsrasterizingdepthgaussian}, VEGS*\cite{hwang2024vegsviewextrapolationurban}, and EVPGS(RaDe-GS) to assess the efficiency of our framework. 
For all three methods, we train each for a total of 30k iterations. Specifically, for EVPGS(RaDe-GS), we follow the coarse-to-fine training process: 20k iterations for pretraining, 1k iterations for the coarse stage, and 9k iterations for the fine stage.
Table~\ref{tab:supp_time} presents the training time of each method across different datasets.
To mitigate artifacts and recover high-frequency details in extrapolated views, EVPGS incorporates several strategies that increase training time compared to the baseline RaDe-GS. While VEGS* is also a GS-based method specifically designed for the EVS problem, EVPGS is both more efficient and achieves superior performance.
As shown in Table~\ref{tab:supp_time} and Table~\co{1} of the main paper, EVPGS effectively balances performance and computational cost.

\begin{table}[h]
    \centering
    \resizebox{1.0\linewidth}{!}{
    \begin{tabular}
    {@{}l@{\,\,}|ccc}
    \textbf{Training Time}
    & DTU & Merchandise3D  & Synthetic-NeRF  \\ \hline
    RaDe-GS~\cite{zhang2024radegsrasterizingdepthgaussian} & $\sim$13.5m & $\sim$15m & $\sim$14.2m \\
VEGS*~\cite{hwang2024vegsviewextrapolationurban} & $\sim$52.5m & $\sim$43.3m & $\sim$35m\\
EVPGS(RaDe-GS) & $\sim$25m & $\sim$26m & $\sim$22.5m
    \end{tabular}
    }
    \caption{\textbf{Computational Cost Comparison of RaDe-GS~\cite{zhang2024radegsrasterizingdepthgaussian}, VEGS*~\cite{hwang2024vegsviewextrapolationurban} and EVPGS(RaDe-GS).}
    We compare the training time of these three methods. Compared to the RaDe-GS baseline, our EVPGS achieves a good balance between performance and computational cost. Additionally, EVPGS is more efficient than VEGS* in handling the EVS problem.
    }
    \label{tab:supp_time}

\end{table}

\subsection{Comparing with Sparse-view GS Methods}

To assess the applicability of sparse-view GS methods in EVS scenarios, we select the state-of-the-art sparse-view GS method MVPGS~\cite{xu2024mvpgsexcavatingmultiviewpriors} as a baseline and integrate it into our EVPGS framework. We uniformly sample 9 training views per scene from our EVS training split (Sec.~\co{4.1} of the main paper) to train MVPGS.
As shown in Table~\ref{tab:supp_sparse}, MVPGS performs poorly with the EVS task, while our EVPGS framework still provides a slight performance improvement.

\begin{table*}[]
    \centering
    \resizebox{1.0\linewidth}{!}{
    \begin{tabular}{@{}l@{\,\,}|ccc|ccc|ccc}
    & \multicolumn{3}{c|}{DTU} & \multicolumn{3}{c|}{Merchandise3D} & \multicolumn{3}{c}{Synthetic-NeRF} \\
    & PSNR $\uparrow$ & SSIM $\uparrow$ & LPIPS $\downarrow$ & PSNR $\uparrow$ & SSIM $\uparrow$ & LPIPS $\downarrow$ & PSNR $\uparrow$ & SSIM $\uparrow$ & LPIPS $\downarrow$\\ \hline
    MVPGS~\cite{xu2024mvpgsexcavatingmultiviewpriors}& 21.032 & 0.7761 & 0.1705 & 11.067 & 0.7505 & 0.2693 & 13.929 & 0.7008 & 0.3613
\\
RaDe-GS~\cite{zhang2024radegsrasterizingdepthgaussian}&   25.778  &  0.8882  & 0.0753  & 23.559   & 0.9134   &  0.0645  & 27.531  & 0.9190   &  0.0530 
\\
\hline
EVPGS(MVPGS)
& 21.471 
& 0.7877
& 0.1643
& 11.133 & 0.7519 & 0.2670 & 13.965 &	0.7017 & 0.3598
\\
EVPGS(RaDe-GS)
&  \textbf{26.488} &
 \textbf{0.8991} &
 \textbf{0.0670} &
 \textbf{25.136}  &
  \textbf{0.9267} &
 \textbf{0.0496}  &
 \textbf{27.849} &
 \textbf{0.9243} &
 \textbf{0.0498} 
    \end{tabular}
    }
    \vspace{-0.1in}
    \caption{\textbf{Quantitative evaluation of our EVPGS framework with MVPGS~\cite{xu2024mvpgsexcavatingmultiviewpriors}.} We evaluate EVPGS integrated with the state-of-the-art sparse-view method MVPGS~\cite{xu2024mvpgsexcavatingmultiviewpriors}. The results indicate that sparse-view methods struggle to handle the EVS scenario effectively. However, our EVPGS framework still provides a slight improvement over MVPGS. 
    }
    \label{tab:supp_sparse}
    \vspace{-0.1in}
\end{table*}

\subsection{Results on real-life dataset}

\begin{table}[h]
    \centering
    \resizebox{1.0\linewidth}{!}{
    \begin{tabular}{@{}l@{\,\,}|ccc}
    & \multicolumn{3}{c}{Mip-NeRF360}  \\
    & PSNR $\uparrow$ & SSIM $\uparrow$ & LPIPS $\downarrow$ \\ \hline
    RaDe-GS & 19.066 & 0.5679 & 0.2253  \\

EVPGS(RaDe-GS)* & \textbf{19.446} & \textbf{0.5757} & \textbf{0.2091} 
    \end{tabular}
    }
    \caption{\textbf{Quantitative evaluation of our EVPGS framework on the Mip-NeRF360~\cite{barron2022mip} dataset.} We compare EVPGS(RaDe-GS) variant $full\: w/o\: mesh$ (denoted as EVPGS(RaDe-GS)*) with the RaDe-GS baseline, demonstrating that our EVPGS framework also achieves improved performance in outdoor scenes.}
    \label{tab:supp_mipnerf360}
\end{table}

\begin{figure}[h]
\centering
\includegraphics[width=1.0\linewidth]{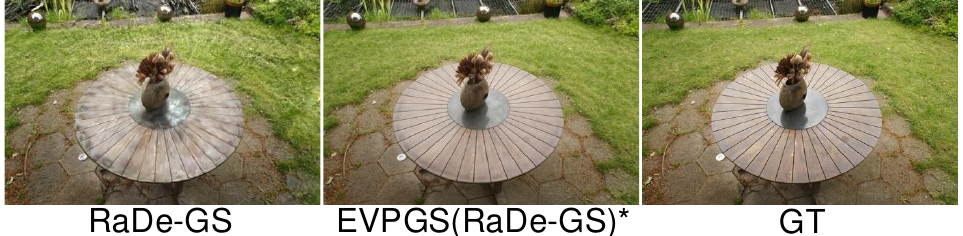}
\vspace{-15pt}
\caption{\textbf{Qualitative Results on Mip-NeRF360~\cite{barron2022mip} dataset.}
The result of EVPGS(RaDe-GS) variant $full\: w/o\: mesh$ (denoted as EVPGS(RaDe-GS)*) show fewer artifacts and preserve more details compared to the RaDe-GS baseline.}
\label{fig:supp_mipnerf360}
\end{figure}

EVPGS is primarily designed for object-centric scenes. To further evaluate its generalizability, we tested it on real-world scenes from the Mip-NeRF360~\cite{barron2022mip} dataset. We created training and testing splits following the EVS setting, resulting in an average pitch angle difference of 17.01° between splits across scenes, compared to 4.10° in the original Mip-NeRF360 splits. Notably, since the reconstructed mesh after the pretraining stage is not available in Mip-NeRF360, we used the EVPGS variant $full\: w/o\: mesh$ (Section~\co{4.3} of the main paper) for evaluation. As shown in Table~\ref{tab:supp_mipnerf360} and Figure~\ref{fig:supp_mipnerf360}, EVPGS achieves a performance boost comparable to that observed in object-centric scenes, demonstrating its potential in scene-level reconstruction.

\subsection{More Qualitative Results on Ablation Study}

In the main paper, we presented the qualitative results of our ablation study in Figure~\co{6}, showcasing two simplified versions of EVPGS: \textit{only fine} and \textit{full w/o ref.}. In addition to Figure~\co{6}, we conduct further qualitative experiments on the DTU~\cite{jensen2014large} and Synthetic-NeRF~\cite{mildenhall2020nerfrepresentingscenesneural} datasets to evaluate two other simplified versions of EVPGS: \textit{only coarse} and \textit{full w/o occ.}. These experiments assess the effectiveness of our fine stage and occlusion-aware module.
We present the results in Figure~\ref{fig:supp_ablation}, which demonstrate that the fine stage (Section~\co{3.3} of the main paper) enhances the reconstruction of fine details, as evidenced by comparing the full EVPGS with \textit{only coarse}. Additionally, the occlusion-aware reprojection strategy (Section~\co{3.3} of the main paper) effectively mitigates image corruption caused by occlusions, as shown by the comparison between the full EVPGS and \textit{full w/o occ.}.

\subsection{More Qualitative Results on EVPGS}
\begin{figure*}
    \centering
    \includegraphics[width=1.0\linewidth]{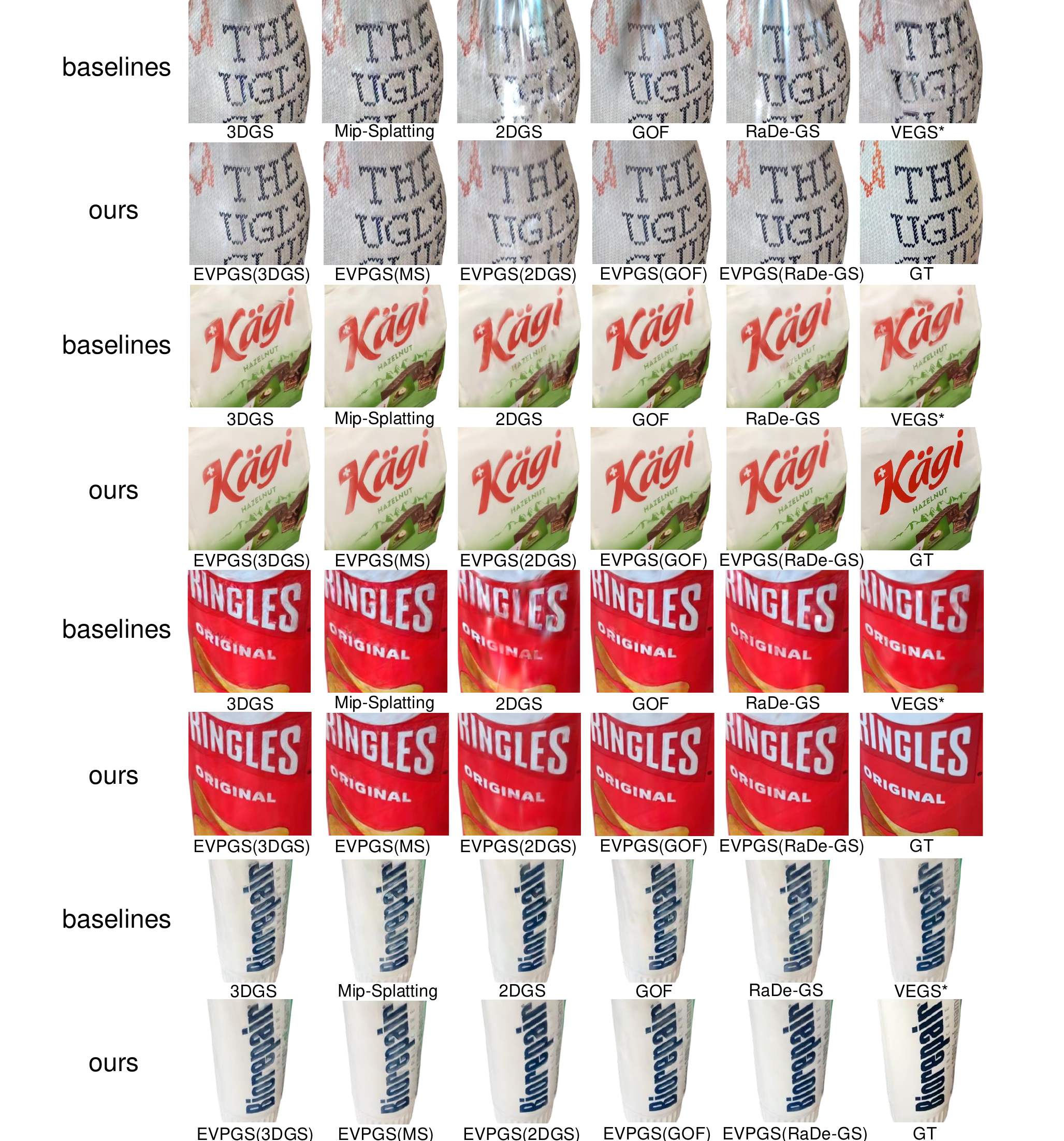}
    \caption{\textbf{More qualitative comparison on our Merchandise3D dataset.}
    }
    \label{fig:supp_Merchandise_1}
\end{figure*}

\begin{figure*}
    \centering
    \includegraphics[width=1.0\linewidth]{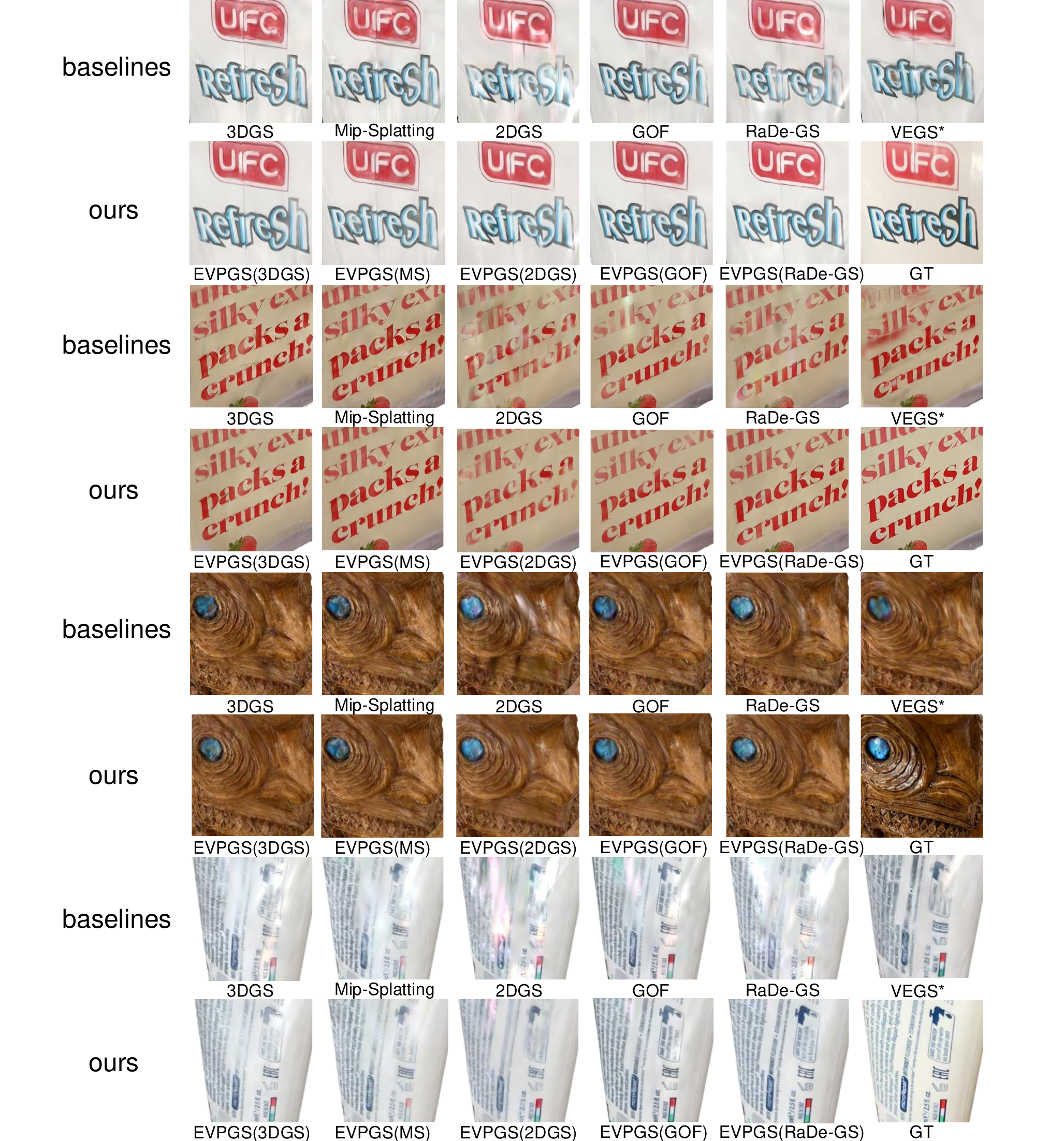}
    \caption{\textbf{More qualitative comparison on our Merchandise3D dataset.}
    }
    \label{fig:supp_Merchandise_2}
\end{figure*}

\begin{figure*}
    \centering
    \includegraphics[width=1.0\linewidth]{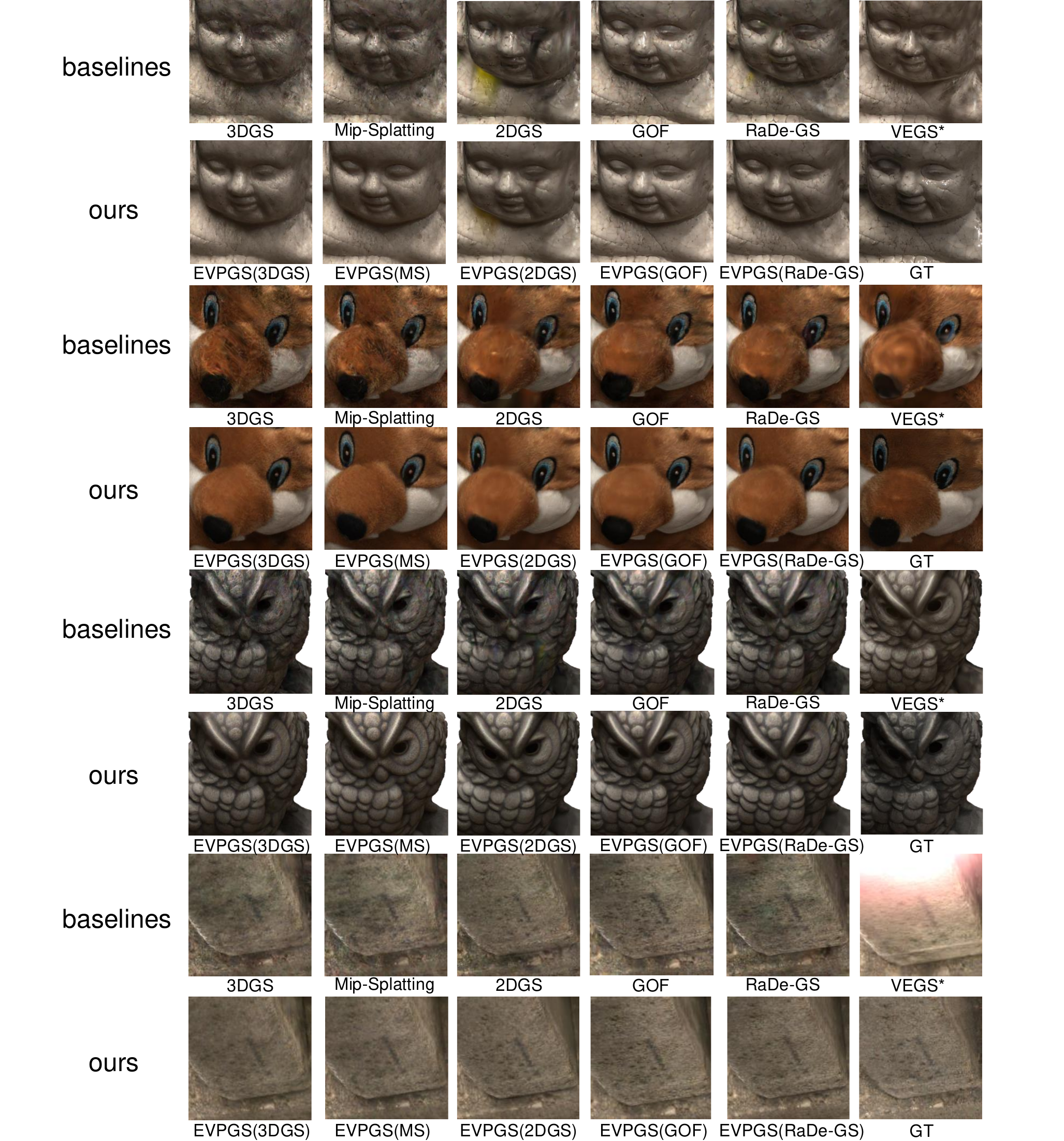}
    \caption{\textbf{More qualitative comparison on the DTU~\cite{jensen2014large} dataset}.  
    }
    \label{fig:supp_dtu_1}
\end{figure*}

\begin{figure*}
    \centering
    \includegraphics[width=1.0\linewidth]{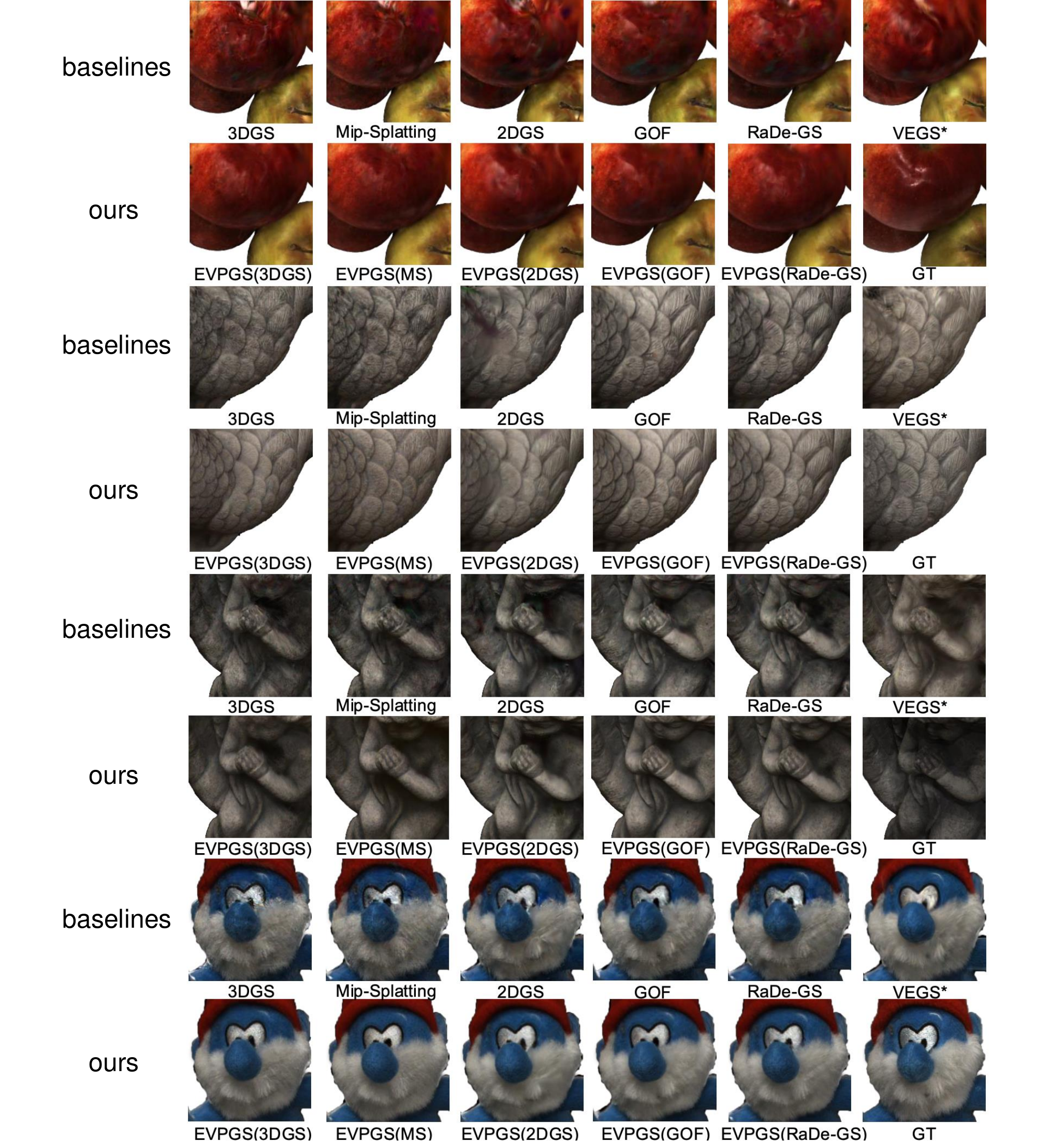}
    \caption{\textbf{More qualitative comparison on the DTU~\cite{jensen2014large} dataset}.  
    }
    \label{fig:supp_dtu_2}
\end{figure*}

\begin{figure*}
    \centering
    \includegraphics[width=1.0\linewidth]{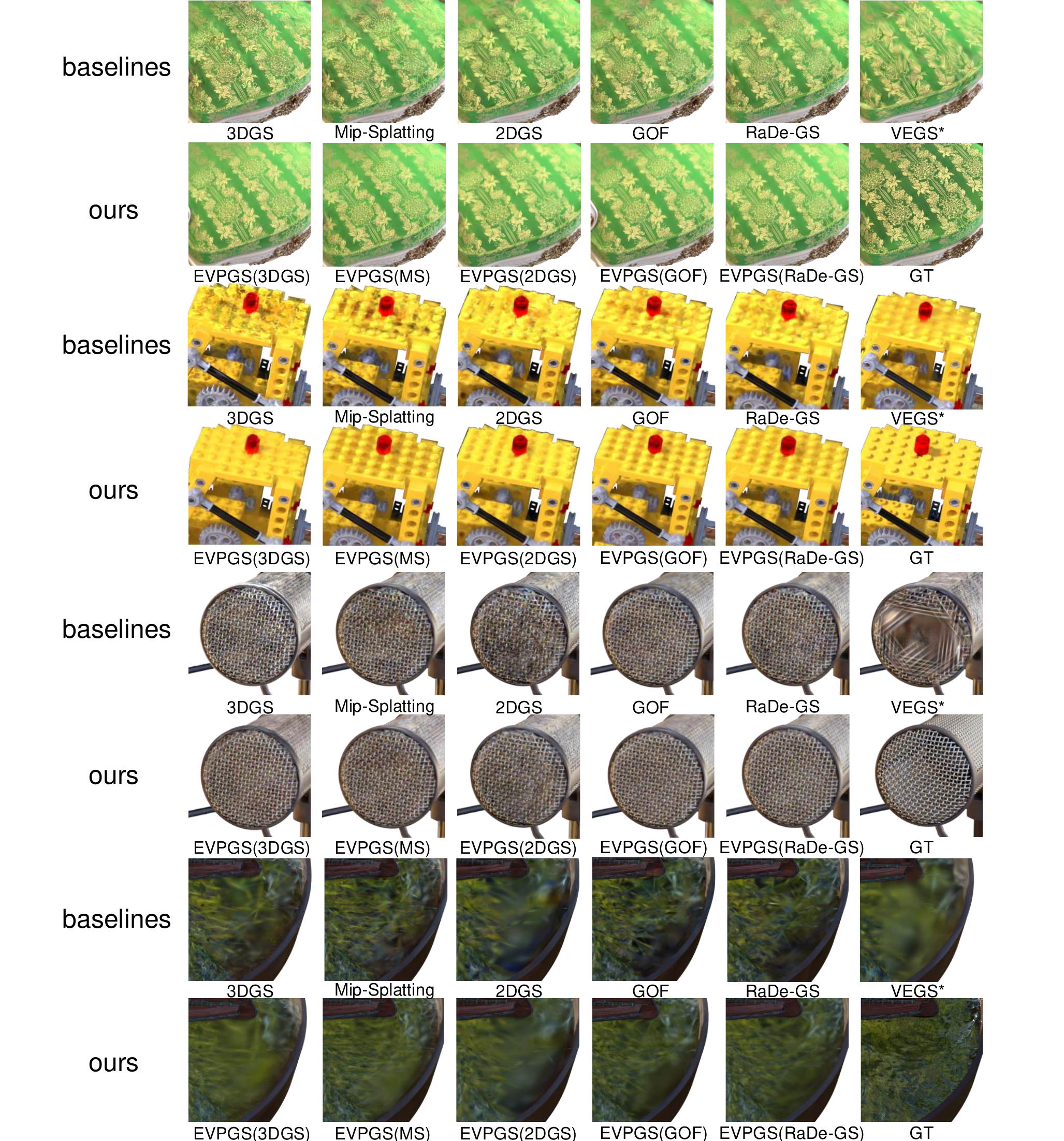}
    \caption{\textbf{More qualitative comparison on the Synthetic-NeRF~\cite{mildenhall2020nerfrepresentingscenesneural} dataset}. 
    }
    \label{fig:supp_synthetic_1}
\end{figure*}

\begin{figure*}
    \centering
    \includegraphics[width=1.0\linewidth]{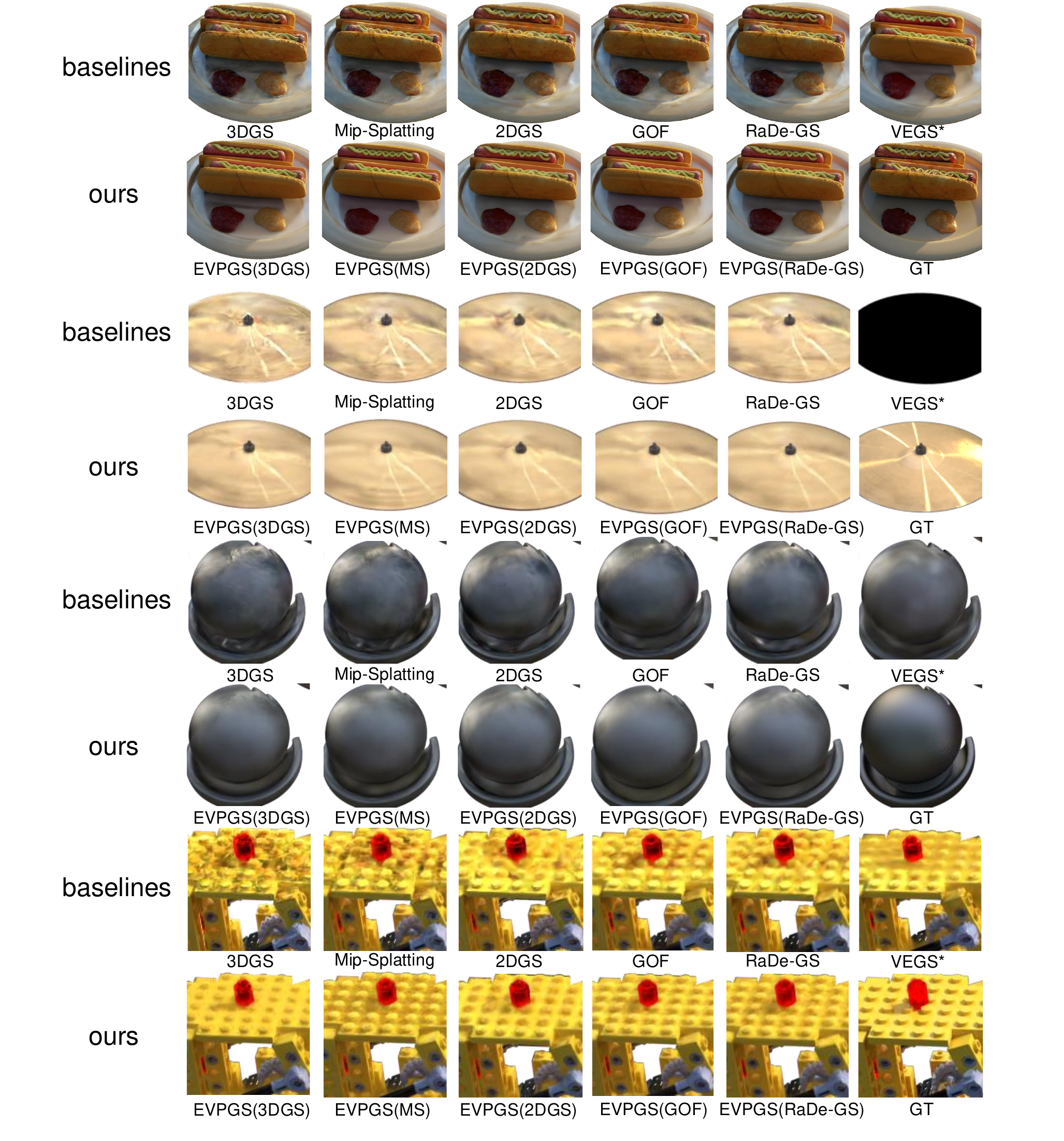}
    \caption{\textbf{More qualitative comparison on the Synthetic-NeRF~\cite{mildenhall2020nerfrepresentingscenesneural} dataset}. 
    }
    \label{fig:supp_synthetic_2}
\end{figure*}

In addition to the qualitative results in Figure~\co{4} and Figure~\co{5} of the main paper where we compared our overall EVPGS with the other methods on the Merchandise3D, DTU~\cite{jensen2014large}, and Synthetic-NeRF~\cite{mildenhall2020nerfrepresentingscenesneural} datasets, we conduct further qualitative comparison on more object instances from these three datasets. We present the additional comparison results in Figure~\ref{fig:supp_Merchandise_1} and Figure~\ref{fig:supp_Merchandise_2} for Merchandise3D, Figure~\ref{fig:supp_dtu_1} and 
Figure~\ref{fig:supp_dtu_2} for DTU~\cite{jensen2014large},  
Figure~\ref{fig:supp_synthetic_1} and Figure~\ref{fig:supp_synthetic_2} for Synthetic-NeRF~\cite{mildenhall2020nerfrepresentingscenesneural}.These results highlight the intricate structures and fine details accurately reconstructed by our EVPGS framework across all three datasets, showcasing the effectiveness of EVPGS in addressing the EVS task.

\section{Application}
\label{sec:application}

Our EVPGS framework enables a practical application for free-view merchandise exhibition, allowing users to effortlessly showcase any object they desire. When creating attractive merchandise videos using the conventional commercial technologies, it often requires a professional photographer to capture the object from a diversity of viewpoints along a set of planned camera paths, which can make the filming process labor-intensive and costly. With EVPGS, users only need to capture a simple circular sequence of images around the object using a smartphone. Then our EVPGS generates high-quality extrapolated views, enabling the creation of engaging videos from various perspectives with minimal effort. EVPGS not only reduces the burden of photography but also ensures display of the object with good realism and high-quality details, providing an efficient and effective solution for merchandise display. 
We provide several merchandise display videos with artistically designed camera paths in our project page. \textbf{We encourage the readers to watch the videos in our \href{https://charley077.github.io/EVPGS_Homepage/}{project page} to gain a better grasp of the capabilities of our EVPGS in real-world applications.}

\end{document}